\documentclass[aps,prb,twocolumn,superscriptaddress,floatfix]{revtex4-1}

\usepackage{graphicx}
\usepackage{amsmath}
\usepackage{amsthm}
\usepackage{dsfont}
\usepackage{color}
\usepackage{amssymb}

\makeatletter
\newcommand*\dashline{\rotatebox[origin=c]{90}{$\dabar@\dabar@\dabar@$}}
\makeatother

\pdfoutput=1

\begin{document}

\title[]{Non-Abelian quasiholes in lattice Moore-Read states and parent Hamiltonians}

\author{Sourav Manna}

\affiliation{Max-Planck-Institut f\"ur Physik komplexer Systeme, D-01187 Dresden, Germany}

\author{Julia Wildeboer}

\affiliation{Department of Physics \& Astronomy, University of Kentucky, Lexington, Kentucky 40506-0055, USA}

\author{Germ\'an Sierra}
\affiliation{Instituto de F{\'i}sica Te{\'o}rica, UAM-CSIC, Madrid, Spain}

\author{Anne E. B. Nielsen}

\affiliation{Max-Planck-Institut f\"ur Physik komplexer Systeme, D-01187 Dresden, Germany}
\affiliation{Max-Planck-Institut f\"ur Quantenoptik, Hans-Kopfermann-Str.\ 1, D-85748 Garching, Germany}
\affiliation{Department of Physics and Astronomy, Aarhus University, DK-8000 Aarhus C, Denmark}

\begin{abstract}
This work concerns Ising quasiholes in Moore-Read type lattice wave functions derived from conformal field theory. 
We commence with constructing Moore-Read type lattice states and then add quasiholes to them. By use of Metropolis Monte Carlo simulations,
we analyze the features of the quasiholes, such as their size, shape, charge, and braiding properties.
The braiding properties, which turn out to be the same as in the continuum Moore-Read state, demonstrate the topological attributes of the Moore-Read lattice states in a direct way.  We also derive parent Hamiltonians for which the states with quasiholes included are ground states. One advantage of these Hamiltonians lies therein that we can now braid the quasiholes just by changing the coupling strengths in the Hamiltonian since the Hamiltonian is a function of the positions of the quasiholes. The methodology exploited in this article can also be used to construct other kinds of lattice fractional quantum Hall models containing quasiholes, for example investigation of Fibonacci quasiholes in lattice Read-Rezayi states. 
\end{abstract}

\maketitle

\section{Introduction}

Strongly correlated quantum many-body systems exhibit a cornucopia of intriguing phenomena that cannot be perceived in conventional materials and are of great importance both for fundamental theoretical studies and experimental points of view. Examples include scenarios such as quantum phase transitions, quantum spin liquids, topological quantum systems and many more. Theoretical progress in this direction is hindered due to high complexity of the many body systems. Numerically the high complexity arises from strong correlations and the exponential growth of the Hilbert space with the system size. Analytical models are therefore very helpful to gain insight.

The fractional quantum Hall effect was a pioneering breakthrough in the context of topological systems \cite{Others49, JKJ21,Moore-Read1,N.Read1}.  This phenomenon unveils an exotic phase of matter \cite{NR8,ZL3,FDMH5,C.Nayak15,C.Nayak19,SDS4,JKJ3,Others44,Others45,Others46,Others47,Others34} and is obtainable at very low temperatures. One of the most important trademarks of fractional quantum Hall states is that they support emergent fractionally charged quasiparticle excitations with non trivial braiding properties \cite{DA1}. While fermions obey Fermi-Dirac statistics and bosons follow Bose-Einstein statistics, these quasiparticles in two dimensional systems follow any-statistics and hence found the  nomenclature as \textit{anyons}\cite{DA1,N.Read1}. In most of the states the statistics is Abelian \cite{RNB1} meaning that under an anyonic winding \cite{C.Nayak18} around each other the wavefunction acquires only a phase factor. More interestingly, if the ground state in a sector is degenerate for fixed anyon positions, an exchange of the anyons corresponds to a unitary matrix transformation and then those anyons exhibit non-Abelian braiding statistics \cite{SDS1,SDS2}. In the present days non-Abelian anyons \cite{Others2,Others43} are drawing much attention both from a theoretical and a practical viewpoint due to their exceptional properties and their potential applications in quantum information especially topologically protected fault tolerant quantum computation \cite{C.Nayak1}. 

Analytical trial wavefunctions are of great importance to understand the fractional quantum Hall effect  \cite{Others49}. One of the promising candidates supporting non-Abelian anyons, the state under consideration here, is the Moore-Read Pfaffian fractional quantum Hall state \cite{IS1,C.Nayak12,JKJ6,JKJ4} at the second Landau level with filling $\nu = \frac{5}{2}$. Construction of fractional quantum Hall states in lattices has recently gained much interest, and the present work is concerned with similarly constructing trial wavefunctions on the lattice of Moore-Read states with anyons. The non-Abelian anyonic excitations \cite{NR5,AS5,Others26} in this state containing an even number of anyons span a degenerate space. The positively charged anyons are called  \textit{quasiholes}. Moore and Read advocated these states containing quasiholes by exploiting conformal field correlators of the underlying Ising conformal field theory (CFT) through the connection between corresponding low-energy effective Chern-Simons gauge theories \cite{C.Nayak2} and the conformal blocks.\\

Moore-Read states on lattices without anyons were constructed \cite{Julia1, Anne4} previously and it was found from entanglement properties of these states that they are in the same topological phase as the Moore-Read states in the continuum. It should hence be possible to also construct quasiholes in the lattice models. The results of this paper show that the wavefunctions with quasiholes can be obtained from those without quasiholes in the same way as for the continuum wavefunctions utilizing CFT. We make a detailed investigation of the quasiholes, including computing their size, shape, charge and braiding statistics. We do explicit computations for the square lattice, but the construction of these states is quite general and the analytical forms are applicable for arbitrary lattices in 2D. We can make detailed investigations, since the analytical form of the wavefunctions allow us to do Monte Carlo simulations that can be done for quite large systems.

It is interesting to ask if these states with quasiholes are the ground states of some Hamiltonians defined on the lattice. We exploit the null field construction of the underlying Ising CFT to construct parent Hamiltonians, supporting an arbitrary even number of quasiholes. It is also found that the quasihole excitations in these states containing $Q$ quasiholes span  a $2^{\frac{Q}{2}-1}$ dimensional degenerate space evidencing the degeneracy as the signature of non-Abelian nature. The Hamiltonians derived are long ranged and contain five-body interactions. 

The structure of the paper is as follows : We fabricate lattice Moore-Read states without and with  quasiholes in Sec.\ \ref{Sec.Anyonic_states}. Next, in Sec.\ \ref{Sec.Anyon_properties}, we analyze the density profile, charge and size of the quasiholes. The braiding statistics are investigated in Sec.\ \ref{Sec.Anyon_braiding}, and parent Hamiltonians are derived for the aforementioned states in Sec.\ \ref{Sec.Hamiltonians}. Sec.\ \ref{Sec.conclusion} concludes the paper. The details of the derivation of the parent Hamiltonians and a sketch of the Metropolis Monte Carlo technique used are given in the Appendices.\\

\section{ Lattice Moore-Read Pfaffian states containing quasiholes from conformal field correlators}\label{Sec.Anyonic_states}

\begin{figure}
	 	\includegraphics[width=0.5\textwidth]{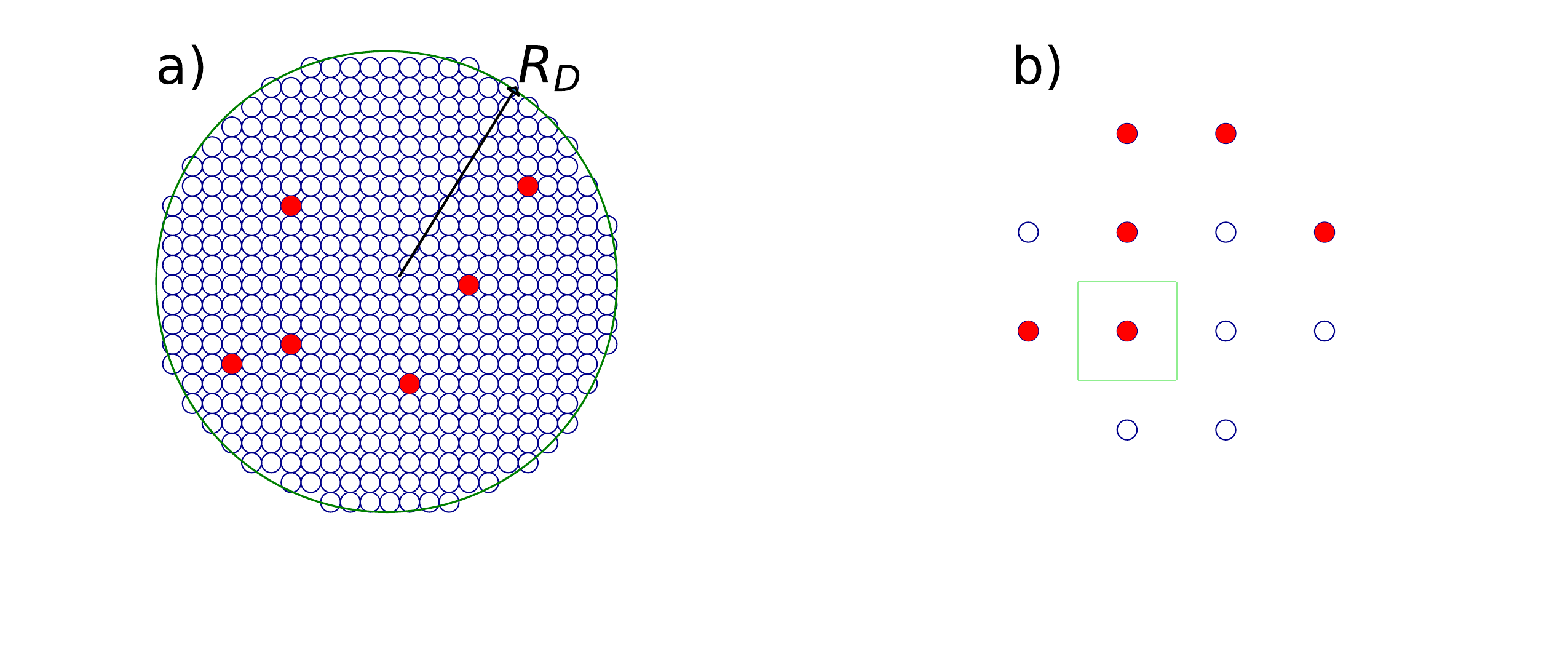}
	 	\caption{In the 2D complex plane the lattice is defined on a disk of radius $R_\mathrm{D}$ as shown in $(a)$. Each site is either empty (blue circles) or singly occupied (red circles). In $(b)$ we mark the area of a lattice site with a square. We illustrate the transformation between the continuum limit $(\eta \longrightarrow 0^+, N \longrightarrow \infty$, for fixed $\eta N$) $(a)$ and the lattice limit $(\eta \longrightarrow 1)$ $(b)$. Note that the lattice filling is $\frac{\eta}{q}$. The interpolation is performed by fixing the number of particles and changing the number of lattice sites per particle between $q$ and infinity.}\label{Fig_lat_cont}
\end{figure}

We introduce and explicitly construct the family of Moore-Read Pfaffian states on lattices hosting quasiholes. Earlier Moore and Read in their pioneering work \cite{Moore-Read1} used CFT in constructing the states for the continuum. We exploit their procedure and construct the states on the lattice. The different members of the family are labeled by the filling fraction $\frac{1}{q}$.

Let us consider an arbitrary lattice in two dimensions with $N$ lattice sites positioned at $z_j$,  $j \in\{1,2,....,N\}$, and the positions of the $Q$ quasiholes are specified by $w_k$ with $k \in\{1,2,....,Q\}$ in the complex plane. We take $a$ to be the area per lattice site and define $\eta = \frac{a}{2\pi}$. This corresponds to that we set the magnetic length to unity. Let us define a local basis at the $j$th site as $|n_j \rangle$, where $n_j$ is the number of particles at site $j$. We have the local basis states as $n_j \in \{0,1\} $ denoting the occupancy of site $j$. Therefore the Hilbert space dimension is $2^N$. The parameter $\eta$ allows us to interpolate between the lattice limit $(\eta \longrightarrow 1)$ and the continuum limit $(\eta \longrightarrow 0^+, N \longrightarrow \infty)$. When doing the interpolation we keep $\eta N$ fixed. As we shall see below, this means that the number of particles per area remains the same, while the number of lattice sites per particle changes from $q$ in the lattice limit to infinite in the continuum limit as displayed in Fig \ref{Fig_lat_cont} for $q = 2$.

To each lattice site, let us associate the vertex operator \cite{Anne4}\\
\begin{equation}\label{Eq_1} 
\mathcal{V}_{n_j}(z_j) =\chi_{n_j}(z_j) \psi(z_j)^{{\Delta_{n_j}}}:e^{i(qn_j-\eta)\phi(z_j)/\sqrt{q}}:
\end{equation}
\begin{equation}\label{Eq_2} 
\chi_{n_j}(z_j) = e^{i\pi(j-1)\eta n_j}
\end{equation}
where $\phi(z_j)$ is the chiral field for the free massless boson of the $U(1)$ CFT with central charge $c = 1$, $\psi(z_j)$ is the holomorphic free Majorana fermion field with conformal dimension $h_\psi = \frac{1}{2}$ of the $c = \frac{1}{2}$ Ising CFT associated with the occupied lattice sites only (since, $\Delta_{n_j} = 1$ iff $n_j = 1$ and $0$ otherwise) and $:....:$ denotes normal ordering. The phase factor $\chi_{n_j}(z_j)$ could be chosen at will. We have taken this particular form since it ensures that the state for $q=1$ is $SU(2)$ invariant when we do not have anyons in the system. Making a different choice of single particle phase factors will not affect the entanglement entropy of the system and hence also not the topological entanglement entropy. The braiding properties, we compute in Sec.\ \ref {Sec.Anyon_braiding}, are also independent of the choice of phase factors.

Now, Moore-Read states can host Ising quasiholes \cite{C.Nayak8}. So, let us introduce\cite{Moore-Read1,German1} the vertex operator\\
\begin{equation}\label{Eq_3}  
W(w_k) = \sigma(w_k) :e^{i p_k \phi(w_k)/\sqrt{q}}:, \quad p_k = \frac{1}{2}
\end{equation}
to each quasihole position $w_k$ where $\sigma(w_k)$ is the holomorphic spin operator of the chiral Ising CFT with conformal dimension $h_\sigma = \frac{1}{16}$ and $\frac{p_k}{q}$ is the charge of the quasihole at $w_k$ (we assume the standard charge of a particle as $ -1$).\\

The wavefunction is defined as 
\begin{equation}\label{Eq_4} 
|\Psi_\alpha\rangle = \frac{1}{C_\alpha} \sum_{n_1,....,n_N}\Psi_\alpha(\vec{w};\vec{z})|n_1,....,n_N\rangle
\end{equation}
where $\Psi_\alpha(\vec{w};\vec{z})$ can be expressed as conformal blocks in the CFT, as was first pointed out in Ref.\ \onlinecite{Moore-Read1} by Moore and Read for the continuum and later extended to lattices \cite{Anne4,Others39}. We have 
\begin{equation}
	C_\alpha^2 = \sum_{n_1,....,n_N} |\Psi_\alpha(\vec{w};\vec{z})|^2
\end{equation}
where $C_\alpha$ is taken to be real. The vectors $\vec{w} = (w_1,....,w_Q)$ and $\vec{z} = (z_1,....,z_N)$ represent the set of quasihole positions and lattice site positions respectively. The underlying C FT in this case is with central charge $c = \frac{1}{2}+1$ where $\frac{1}{2} $ and $1$ describe the Ising contribution, i.e.\ the
Pfaffian part,  and the Jastrow factor of the wavefunction respectively. One could write the correlator of the above mentioned operators as the product of the two aforementioned CFT theories as
\begin{equation}\label{Eq_5a} 
\begin{split}
 \Psi_\alpha  (\vec{w};\vec{z}) & = \langle 0|\prod_{k=1}^{Q}W(w_k)\prod_{j=1}^{N}\mathcal{V}_{n_j}(z_j)|0\rangle_\alpha
\\& = \mathcal{I}_\alpha \times \mathcal{J}
\end{split}
\end{equation}
with
\begin{equation}\label{Eq_5b}
	\mathcal{I}_\alpha =  \langle 0|\prod_{k=1}^{Q}\prod_{j=1}^{N}  \sigma(w_k)   \psi(z_j)^{{\Delta_{n_j}}}|0\rangle_\alpha
\end{equation}
and
\begin{equation}\label{Eq_5c}
	\mathcal{J} = \langle 0|\prod_{k=1}^{Q}\prod_{j=1}^{N}  :e^{i p_k \phi(w_k)/\sqrt{q}}: :e^{i(qn_j-\eta)\phi(z_j)/\sqrt{q}}:|0\rangle
\end{equation}
where $\langle 0|....|0 \rangle$ stands for the vacuum expectation value in the CFT, $\mathcal{I}_\alpha$ stands for the Ising contribution  and $\mathcal{J}$ for the Jastrow contribution coming from the $c=1$ bosonic sector. 

The holomorphic spin operators $\sigma$ of the Ising CFT have  many conformal blocks depending on the number of quasiholes considered. The total number of different labels of the conformal blocks which we denote by the vector $\alpha$ in \eqref{Eq_4} is $2^{\frac{Q}{2}-1}$. Hence, the wavefunctions $\Psi_\alpha$ represent the degenerate set of wavefunctions for fixed quasihole positions and thereby forming the basis for their non-Abelian statistics.

The fusion channel of the Ising fields $\sigma(w_{2i-1})$ and $\sigma(w_{2i})$ is specified by the $i$th entry of the vector $\alpha$. If $\alpha_i = 0 $ or $ \alpha_i = 1$, those fuse to the identity (I) or Majorana fermion field ($\psi$) respectively by following the fusion rule, $\sigma \times \sigma = I + \psi$. For the correlator to be non-zero, all the fields must be fused to the identity \cite{German1} by following the non trivial Ising fusion algebra \cite{C.Nayak8} as 
\begin{equation}\label{Eq_6} 
\begin{split}
 \psi \times \psi = I ;  \quad \psi \times \sigma = \sigma ;  \quad  \sigma \times \sigma = I + \psi
\end{split}	
\end{equation}
 Now, the factors coming from the $c = 1$ CFT theory are the same for an arbitrary even number of quasiholes but the $c = \frac{1}{2}$ CFT gives rise to different terms depending on the number of quasiholes in the state \cite{German1}. We derive the wavefunctions with zero, two and four quasiholes in details below. In the following we shall use the notation that $(z^\prime_1, \cdots, z^\prime_M)$ are the positions of the occupied lattice sites where we denote $M$ to be the number of particles.

\subsection{The boson ($c = 1$ CFT) part of the wavefunction for an arbitrary even number of quasiholes}

Explicit evaluation of the correlator in \eqref{Eq_5c} by standard methods \cite{Others40} results in the following expression

\begin{equation}\label{Eq_7c} 
	\begin{split}
	\mathcal{J} &= \delta_n \prod_{i<j}(z_{i}-z_{j})^{qn_{i}n_{j}}	
	\prod_{i\neq j}(z_{i}-z_{j})^{-\eta n_{i}}\\&
	\times\prod_{i<j}(z_{i}-z_{j})^{\eta^2/q}	
	\prod_{i<j}(w_i-w_j)^{p_ip_j/q}\\&
	\times\prod_{i,j}(w_i-z_j)^{p_in_j}
	\prod_{i,j}(w_i-z_j)^{-\eta p_i/q}
	\end{split}
\end{equation}
where $\delta_n = 1$ iff the total number of particles 

\begin{equation}
	M = \sum_{j=1}^{N} n_j = (\eta N - \sum_{k=1}^{Q}p_k)/q
\end{equation}
and otherwise $\delta_n = 0$. 

In this model the background charge is included by the operators in Eq \eqref{Eq_1}. The lattice filling fraction is defined to be $M/N$ and in the absence of quasiholes (i.e.\ $Q = 0$) and for $\eta = 1$ this is equal to the Landau level filling fraction $1/q$ in the fractional quantum Hall effect.

The $\prod_{i<j}^{N}(z_{i}-z_{j})^{qn_{i}n_{j}}$ factor in \eqref{Eq_7c} can be interpreted as the attachment of flux $q$ to each particle and the $\prod_{i\neq j}^{N}(z_{i}-z_{j})^{-\eta n_{i}}$ factor represents the background charge in the lattice.

The construction in \eqref{Eq_5a} resembles closely the continuum limit where the wavefunctions are generally expressed in the basis spanned by the position of the particles. That means $(z_1^\prime,....,z_M^\prime)$ form a basis of the Hilbert space and the background charge is supported by the Gaussian factors. Hence the charge neutrality is ensured. Let us take the states \eqref{Eq_5a} on a disk $\mathrm{D}$ of radius $R_\mathrm{D} \longrightarrow \infty$ and $N\longrightarrow \infty$ with a fixed number of particles $M$ to reach the continuum limit. In this limit, we approach the usual Gaussian factors \cite{Anne5} as it can be shown that
\begin{equation}\label{Eq_8} 
\begin{split}
&\prod_{j\neq l}(z_{l}-z_{j})^{-\eta n_{l}} \propto e^{-i\sum_{l}^{N}g_l} e^{-\frac{1}{4}\frac{2\pi \eta}{a}\sum_{l}^{N}n_l |z_l|^2}\\&
\prod_{l,j}(w_l-z_j)^{-p_l/q} \propto e^{-i\sum_{l}^{Q}f_l} e^{-\frac{1}{4}\frac{2\pi}{a}\sum_{l}^{Q}\frac{p_l}{q} |w_l|^2}
\end{split}
\end{equation}
where $g_l = $ Im $[\eta\sum_{j(\neq l)}^{N}n_l \ln(z_l-z_j)]$ and $f_l = $ Im $[\frac{1}{q}\sum_{j}^{N}p_l \ln(w_l-z_j)]$ are real numbers giving rise to the phase factors (overall gauge factors) that can be transformed away if needed. The phase factors do not hamper properties like particle-particle correlations and the entanglement entropy of the state.

\subsection{Wavefunction without quasiholes}

When there are no quasiholes,

\begin{equation}\label{Eq_no_qh_1} 
\begin{split}
\mathcal{I}_\alpha  = \langle \psi(z_1^\prime)....\psi(z_M^\prime) \rangle_\alpha 
=\text{Pf} \bigg(\frac{1}{z_i^\prime-z_j^\prime}\bigg)
\end{split}
\end{equation}
where
$M$ is even and 'Pf' stands for the 'Pfaffian'. The Pfaffian is antisymmetric, so the states in \eqref{Eq_5a} are bosonic (fermionic) for $q$ odd (even).

\subsection{The Ising ($c = \frac{1}{2}$ CFT) part of the wavefunction for two quasiholes}

We now consider the case of two quasiholes. There are two independent possibilities depending on the fusion channel of the two Ising spins $\sigma$ as mentioned in \eqref{Eq_6}.   When two $\sigma$ fields fuse to the identity $(I)$ (Majorana $(\psi)$), we have an even (odd) number of Majorana fields in \eqref{Eq_5b} i.e.\ an even (odd) number of particles $M$. We are interested in the fusion channel of output identity $(I)$ since the expression for the correlator in this case is simpler. As there exists only a single generator of the braid group, the two quasiholes behave as if they are Abelian. Later on we shall see that the presence of four quasiholes will show that they are really non-Abelian.

The exact form of the conformal blocks can be achieved through bosonization. Explicit evaluation gives rise to the factors \cite{German1} 
\begin{equation}\label{Eq_9} 
\begin{split}
\mathcal{I}_\alpha =   2^{-\frac{M}{2}} (w_1-w_2)^{-\frac{1}{8}} \times \prod_{i,j}(w_i-{z^\prime_j})^{-\frac{1}{2}}\times \text{Pf}(A)
\end{split}
\end{equation}
where
\begin{equation}\label{Eq_10} 
A_{ij} = \frac{(z_i^\prime - w_1)(z_j^\prime - w_2)+(z_i^\prime - w_2)(z_j^\prime - w_1)}{(z_i^\prime-z_j^\prime)}
\end{equation} 
and $M$ is even. We have here $\alpha = I$.

\subsection{The Ising ($c = \frac{1}{2}$ CFT) part of the wavefunction for four quasiholes}
 \begin{figure*}
	 \includegraphics[width=0.5\textwidth]{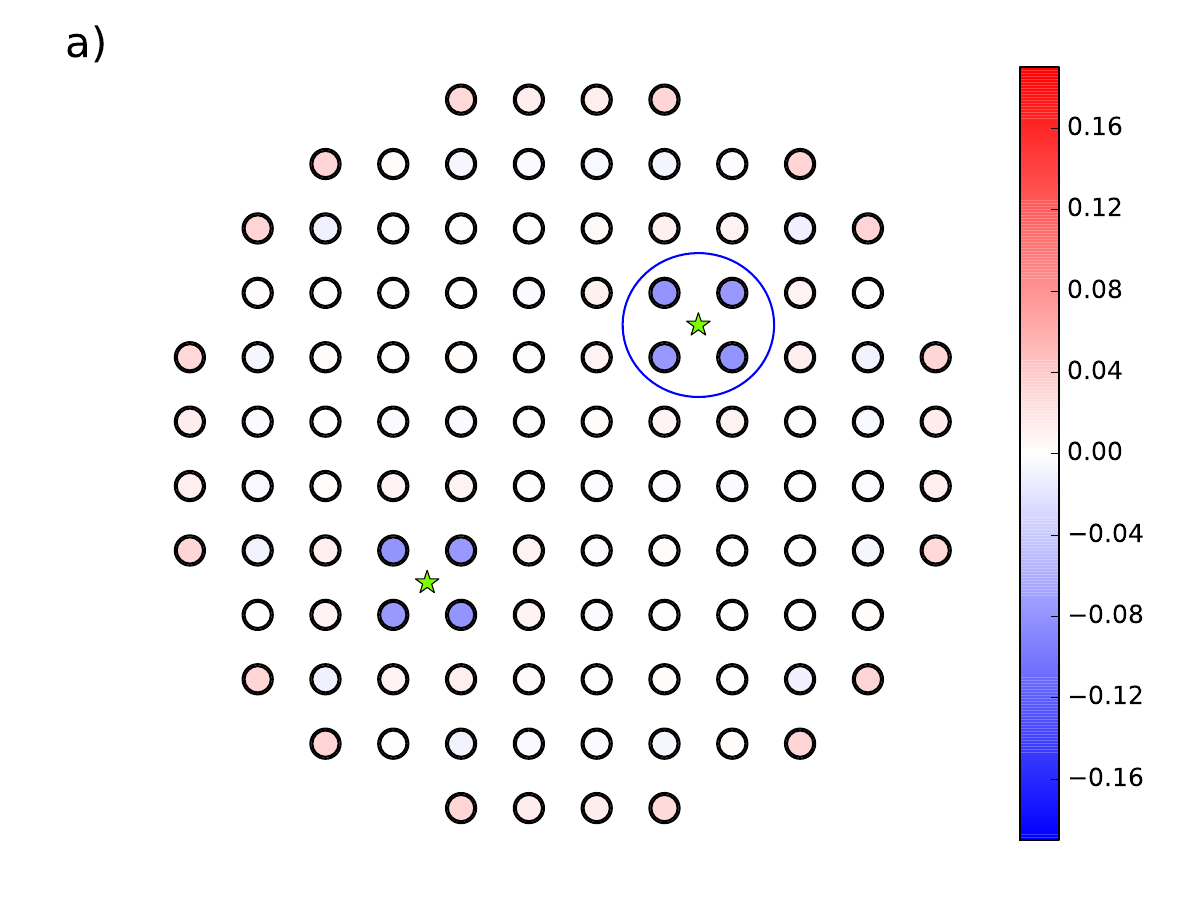}\hfill
         \includegraphics[width=0.5\textwidth]{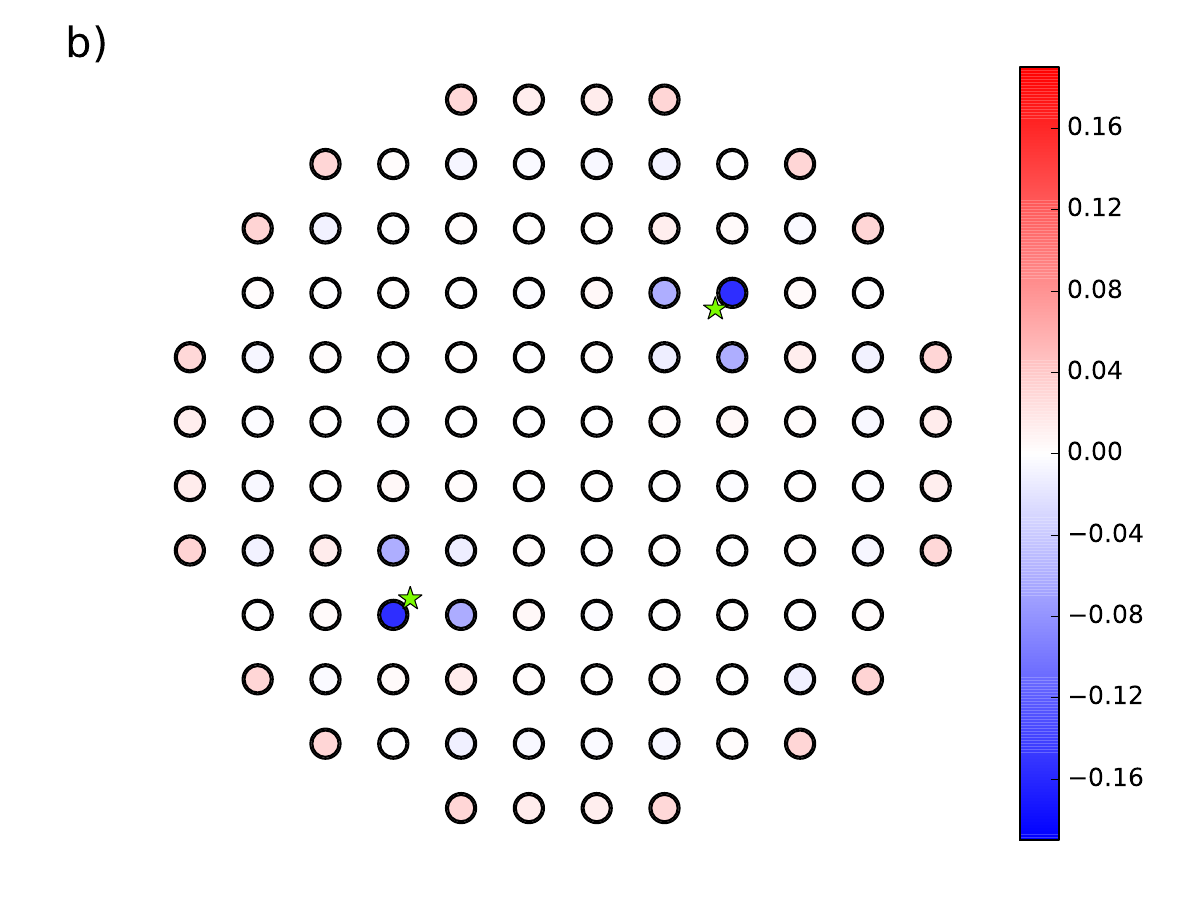}
	 \includegraphics[width=0.5\textwidth]{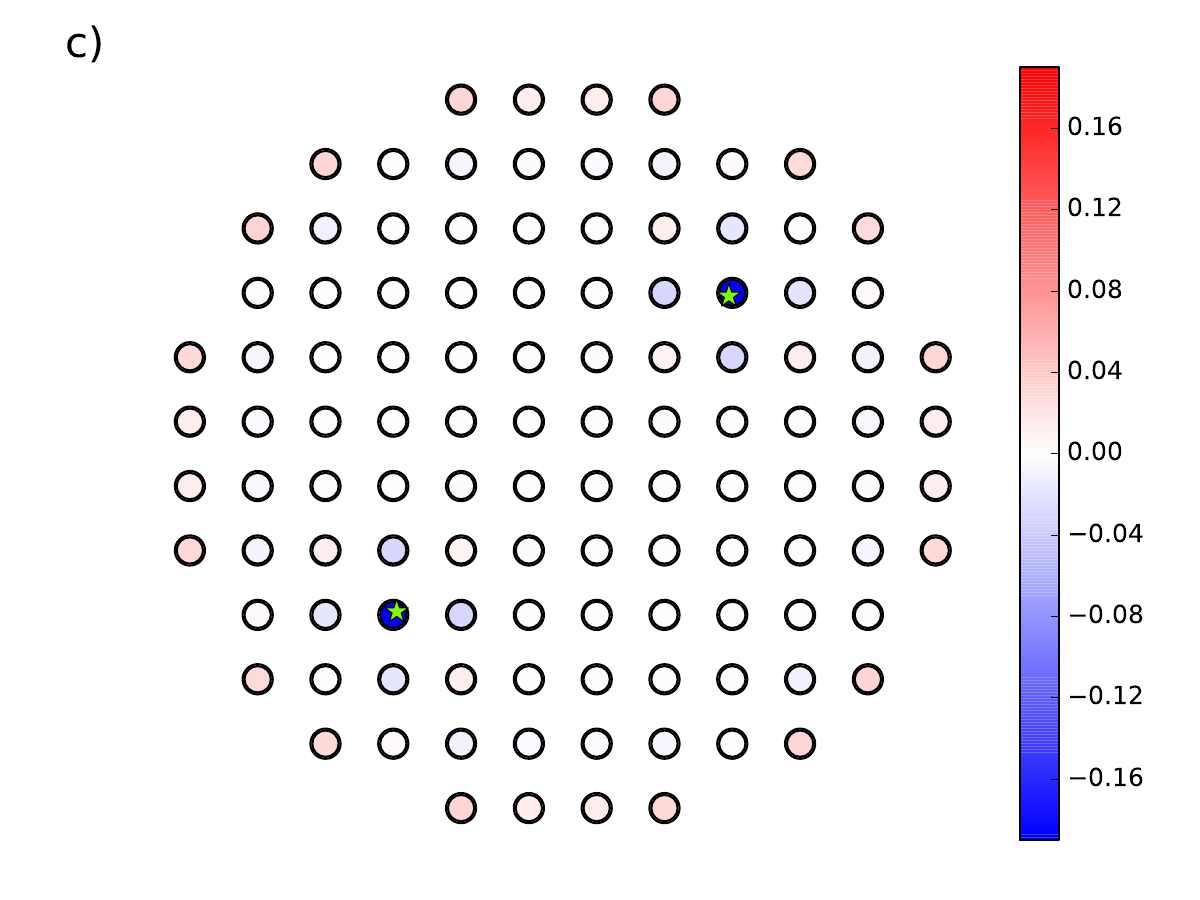}\hfill
         \includegraphics[width=0.5\textwidth]{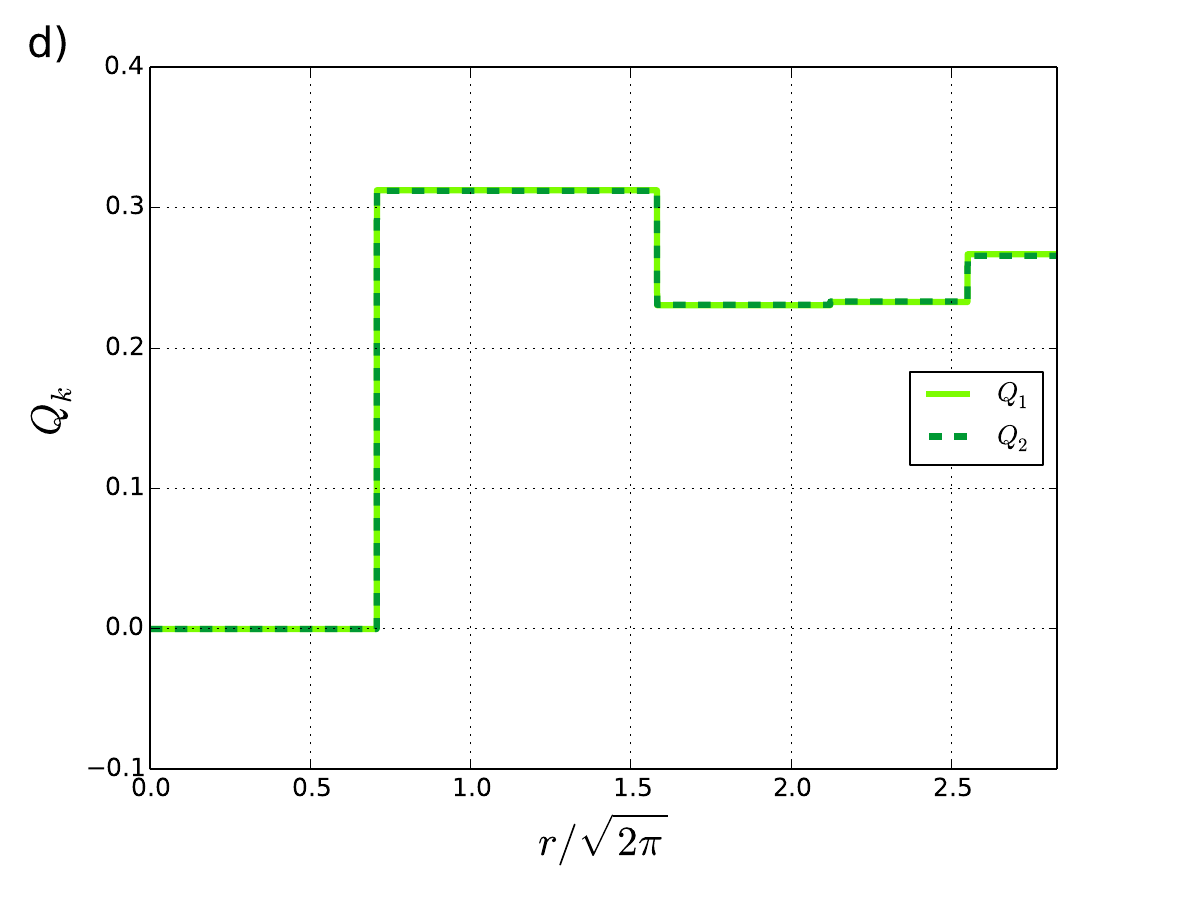}
   \caption{ We mark the lattice sites by the circles and quasiholes by stars. $(a)-(c)$ show the difference between the lattice densities in the presence and absence of the quasiholes in the states \eqref{Eq_5a} i.e.\ $\rho(z_i)= \langle n(z_i) \rangle_{Q \neq 0} - \langle n(z_i) \rangle_{Q = 0}$ with the values represented by the colorbar with an errorbar of size $\sim 10^{-4}$ arising from the Monte Carlo simulation. $q = 2$ and the number of lattice sites is $N=112$. We take the number of particles to be $M = 56$. In $(a)$, the quasiholes are placed exactly in the middle of the plaquette. It turns out that they are screened well and localized with radii of a few lattice constants. The circle shows the radius of an Ising quasihole in the continuum as computed in Ref \onlinecite{BAB14}. If the quasiholes approach the lattice sites, no singularity appears as depicted in $(b)$ and $(c)$. The excess charge (see \eqref{Eq_14}) of the quasihole $Q_{k},\ k \in \{1,2\}$ is computed from $(a)$ and plotted in $(d)$ as a function of the radial distance $r/2\pi$. The quasihole positions are symmetric with respect to a $\pi$ rotation of the lattice. Therefore the two plots are on top of each other.  The charges are approaching the expected value $\simeq 0.25$ for large $r$. The colors on the edges appear because we place a charge at infinity in the state with quasiholes as explained in the text}\label{Fig_1}
  \end{figure*}
To achieve non-Abelian statistics, multiple degenerate states for fixed quasihole positions are necessary. The case of four quasiholes is the simplest one to unveil this behavior \cite{JKJ6}. There are two generators for the four quasihole braid group, giving rise to two different braids, which do not commute with each other and hence forming non-Abelian statistics. Here, $Q = 4$ and hence there are two conformal blocks for \eqref{Eq_5a} which give rise to the degeneracy. We denote the conformal block indices as $ m_I = 0$ and $m_\psi = 1$.

Comprehensive evaluation of the correlator in \eqref{Eq_5b} for two different fusion channels gives \cite{German1}
\begin{equation}\label{Eq_11} 
\begin{split}
\mathcal{I}_\alpha &= 2^{-\frac{M+1}{2}}
(w_1-w_2)^{-\frac{1}{8}} (w_3-w_4)^{-\frac{1}{8}} \times \\& \prod_{i,j}(w_i-z_j^\prime)^{-\frac{1}{2}}
\bigg ((1-x)^{\frac{1}{4}} + \frac{(-1)^{m_\alpha}}{(1-x)^{\frac{1}{4}}}\bigg )^{-\frac{1}{2}} \times \\&
\bigg( (1-x)^{\frac{1}{4}} \Phi_{(13)(24)} + (-1)^{m_\alpha} (1-x)^{-\frac{1}{4}} \Phi_{(14)(23)} \bigg)
\end{split}
\end{equation}
with

\begin{eqnarray}\label{Eq_12} 
& &\Phi_{(k_1k_2)(k_3k_4)} = \nonumber \\
& &\text{Pf}\bigg(  \frac{(w_{k_1}-z_i^\prime)(w_{k_2}-z_i^\prime)(w_{k_3}-z_j^\prime)(w_{k_4}-z_j^\prime) + (i \longleftrightarrow j)}{(z_i^\prime-z_j^\prime)} \bigg) \nonumber \\
& &
\end{eqnarray}
and
\begin{equation}\label{Eq_x} 
\begin{split}
x = \frac{(w_1-w_2)(w_3-w_4)}{(w_1-w_4)(w_3-w_2)}
\end{split}
\end{equation}
where $M$ is even and $x$ is the anharmonic ratio. In $\Phi_{(k_1k_2)(k_3k_4)}$, the quasiholes are labeled by $k_i$ and we have $\alpha \in \{I, \psi\}$.\\

 \section{Density profile and Charge of the quasiholes}\label{Sec.Anyon_properties}

 \begin{figure*}
 	\includegraphics[width=0.5\textwidth]{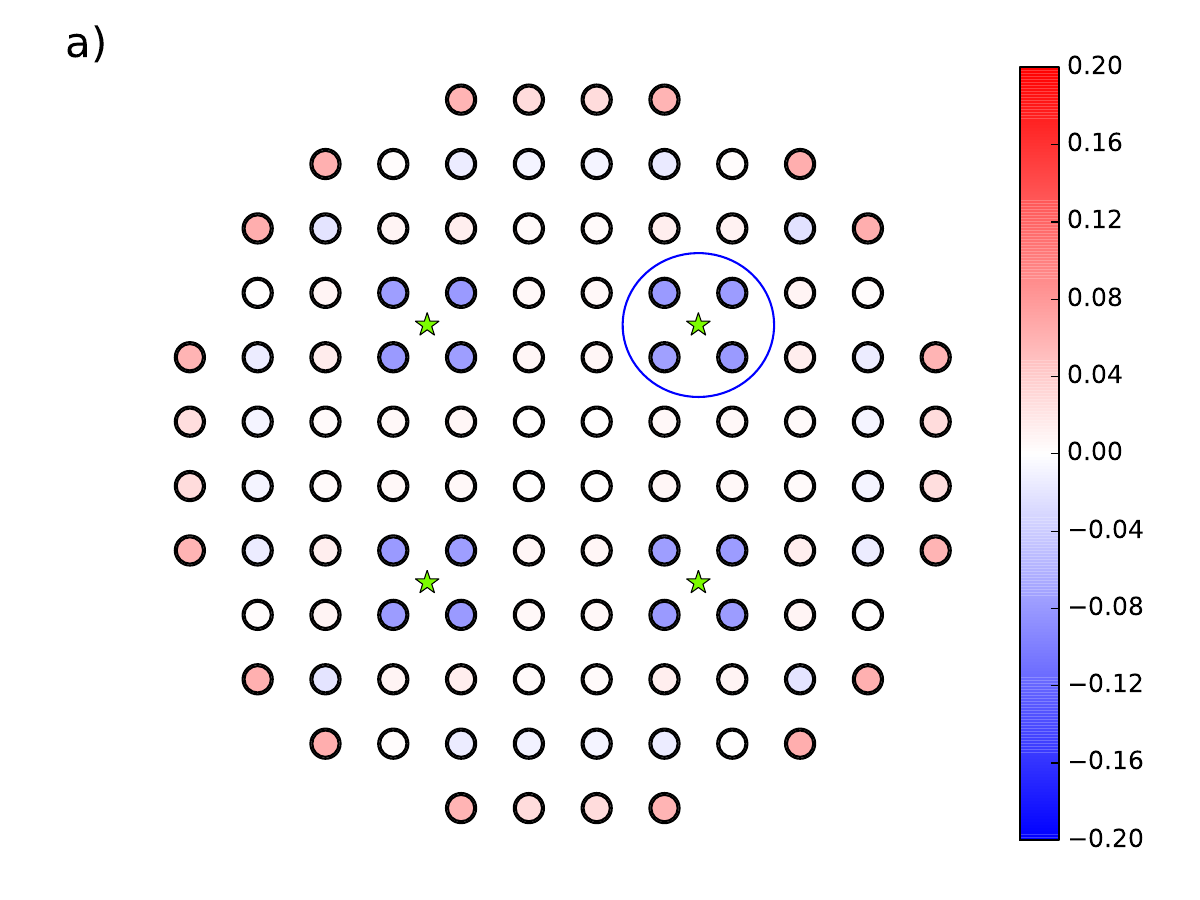}\hfill
 	\includegraphics[width=0.5\textwidth]{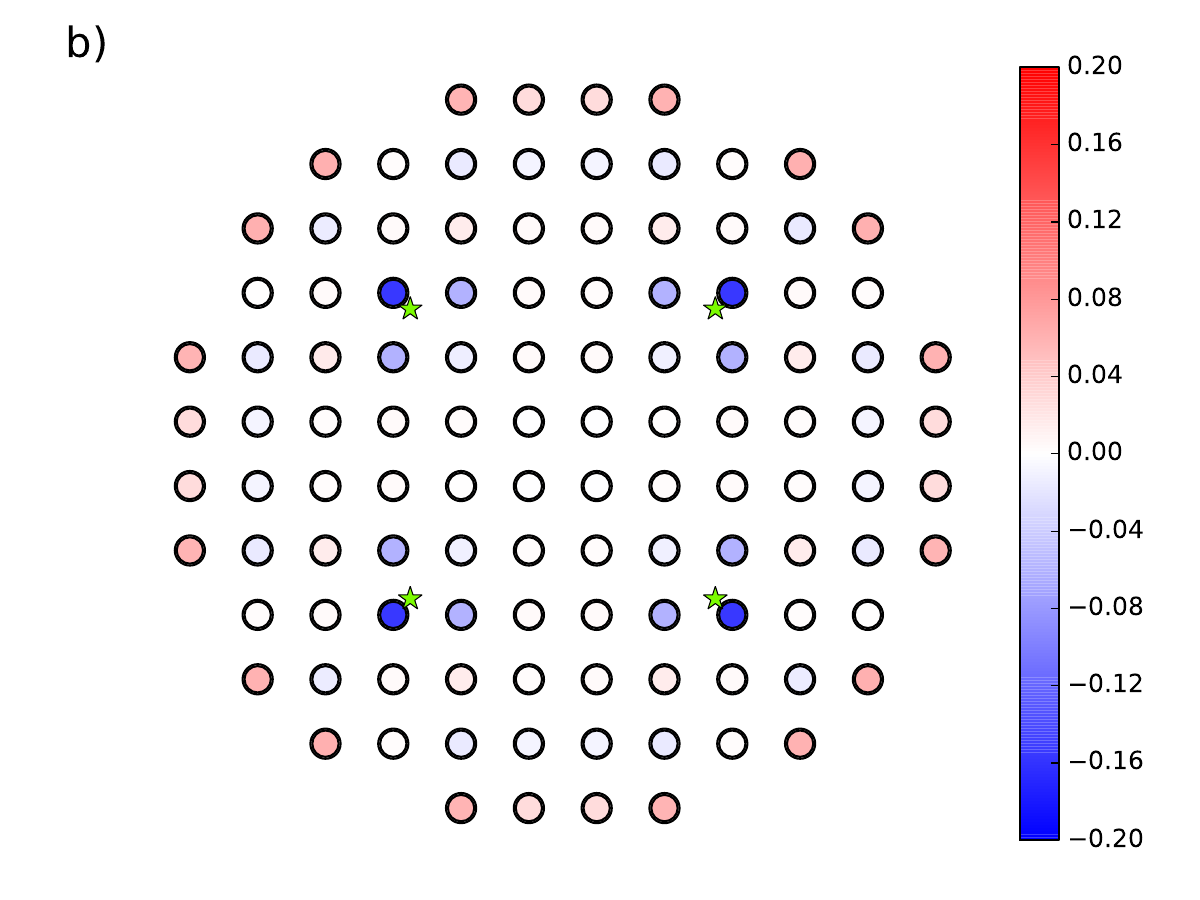}\\
 	\includegraphics[width=0.5\textwidth]{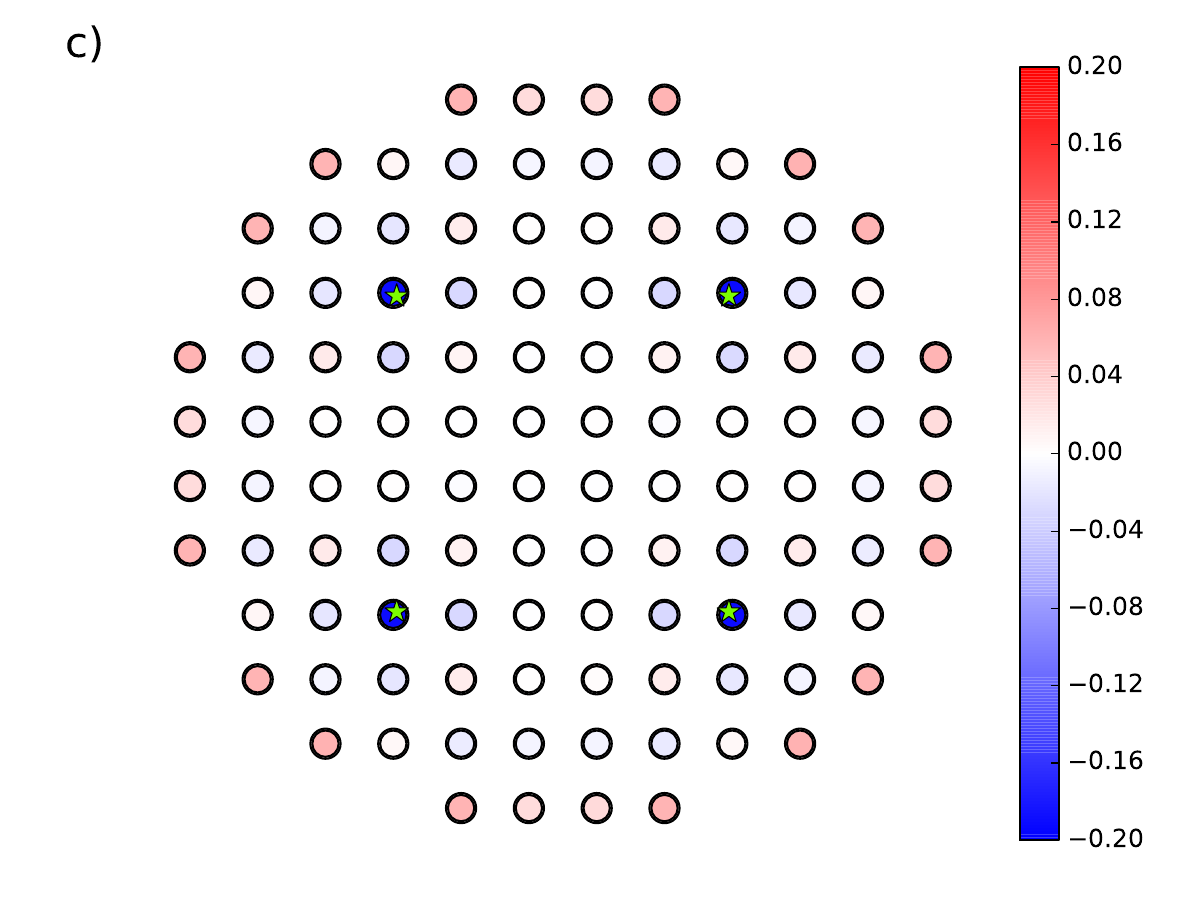}\hfill
 	\includegraphics[width=0.5\textwidth]{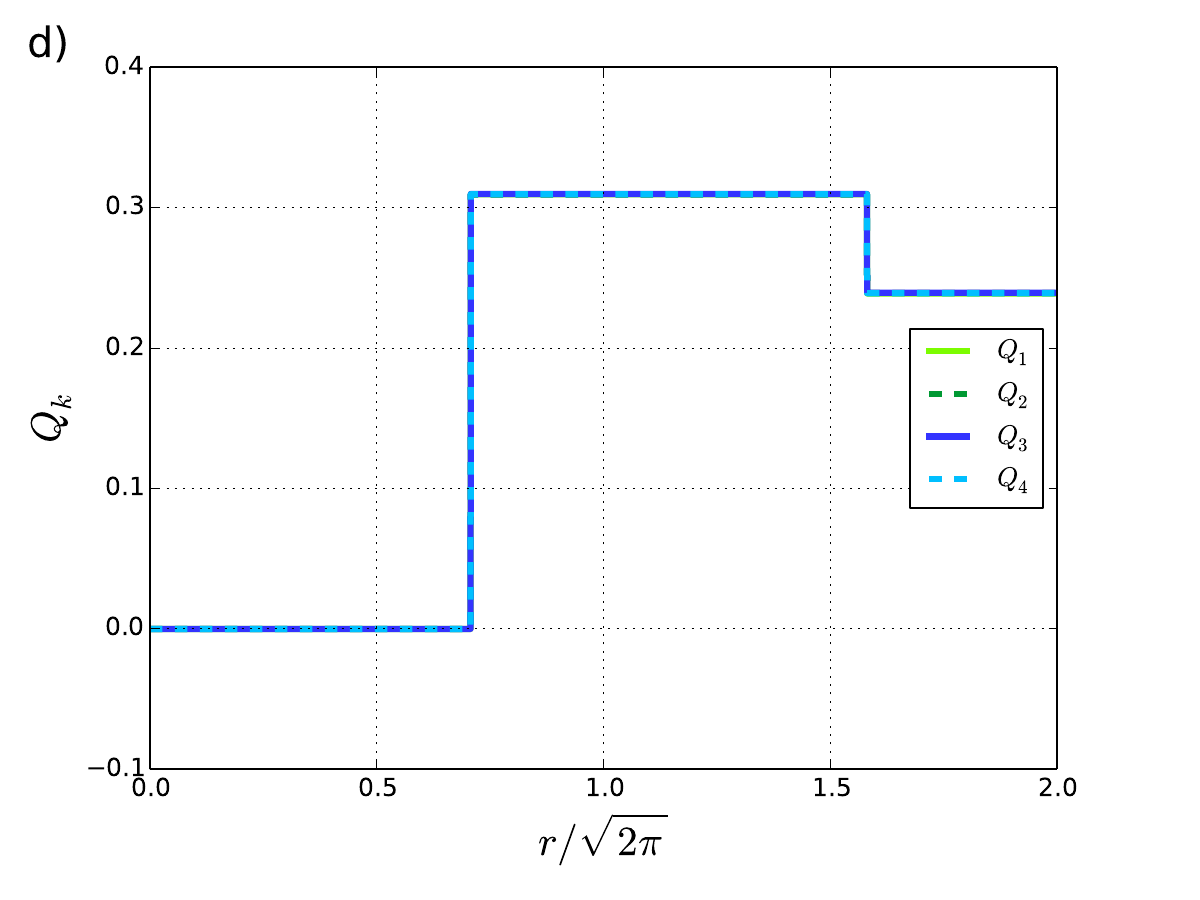}
 	\caption{We mark the lattice sites by the circles and quasiholes by stars. $(a)-(c)$ show the difference between the lattice densities in the presence and absence of the quasiholes in the states \eqref{Eq_5a} i.e.\ $\rho(z_i)= \langle n(z_i) \rangle_{Q \neq 0} - \langle n(z_i) \rangle_{Q = 0}$ with the values represented by the colorbar with an errorbar of size $\sim 10^{-4}$ arising from the Monte Carlo simulation. $q = 2$ and the number of lattice sites is $N=112$. We take the number of particles to be $M = 56$. In $(a)$, the quasiholes are placed exactly in the middle of the plaquette. It turns out that they are screened well and localized with radii of a few lattice constants. The circle shows the radius of an Ising quasihole in the continuum as computed in Ref \onlinecite{BAB14}. If the quasiholes approach the lattice sites, no singularity appears as depicted in $(b)$ and $(c)$. The excess charge (see \eqref{Eq_14}) of the quasihole $Q_{k},\ k \in \{1,2,3,4\}$ is computed from $(a)$ and plotted in $(d)$ as a function of the radial distance $r/2\pi$. The quasihole positions are symmetric with respect to a $\pi/2$ rotation of the lattice. Therefore the four plots are on top of each other.  The charges are approaching the expected value $\simeq 0.25$ for large $r$. The colors on the edges appear because we place a charge at infinity in the state with quasiholes as explained in the text }\label{Fig_2}
 \end{figure*}

We next investigate important properties of the quasiholes. In this section we investigate how the quasiholes influence the density of the particles in the lattice sites, what amount of charges are carried by the quasiholes and how far they extend in the lattice system. We use Metropolis Monte Carlo simulation to research the above mentioned properties for two and four quasiholes. In the numerical computations in this section and the next, we take $q = 2$.\\

\textit{Density profile. -} We define the lattice density of the $i$th lattice site for any state $\Phi$ to be $\langle n(z_i) \rangle = \langle \Phi | n(z_i) | \Phi \rangle$. The density profile of the quasiholes is evaluated as \cite{Anne5}
\begin{equation}\label{Eq_13}
\rho(z_i)= \langle n(z_i) \rangle_{Q \neq 0} - \langle n(z_i) \rangle_{Q = 0}
\end{equation}
where $ \langle n(z_i) \rangle_{Q \neq 0 }$ and $ \langle n(z_i) \rangle_{Q = 0 }$ are the densities of the $i$th lattice site in the presence and absence of the quasiholes in the states respectively. Since, we have the restriction of $\sum_{j=1}^{N} n_j = (\eta N - \sum_{k=1}^{Q}p_k)/q$, it is the case that the insertion of a quasihole leads to the decrement of the total number of particles in the system by $\frac{p_k}{q}$.

Now,  we require the Pfaffian factors in both the wavefunctions containing the quasiholes (Eq.\ \eqref{Eq_5a}) and without quasiholes (Eq.\ \eqref{Eq_no_qh_1}) to be non-zero. Thereby, it  is necessary to set the number of particles $M$ to be even in both the cases. Now, by inspecting the $\delta_n$ factor in \eqref{Eq_5a} it is found that  we can not fulfill this condition simultaneously if we choose the same $\eta$ value for both the cases.  We overcome this problem by inserting an extra charge $\mathcal{P}/q$ at infinity. By choosing appropriate values of this charge we can use the same $\eta$ and make $M$ even in both the cases. Let us incorporate the operator $\Xi_\mathcal{P}(\infty) =   \       :e^{i\frac{\mathcal{P}}{\sqrt{q}}\phi(\infty)}:$ of charge $\frac{\mathcal{P}}{q}$, placed at infinity in the correlator of the wavefunction \eqref{Eq_5a}. Then the wavefunction becomes
\begin{equation}\label{Eq_15} 
\begin{split}
& \Psi_\alpha  (\vec{w};\vec{z}) [\mathcal{P}( \xi \rightarrow \infty)]\propto  \delta^\prime_n \ \mathcal{I}_\alpha 
\prod_{i<j}(z_{i}-z_{j})^{qn_{i}n_{j}}	\\&\times 
\prod_{i\neq j}(z_{i}-z_{j})^{-\eta n_{i}}
\prod_{j}(\xi-z_{j})^{\mathcal{P}n_{j}} \prod_{i,j}(w_i-z_j)^{p_in_j},\\&
\propto  \delta^\prime_n \ \mathcal{I}_\alpha  \ \xi^{\mathcal{P}(N-\mathcal{P})/q}
 \prod_{i<j}(z_{i}-z_{j})^{qn_{i}n_{j}}	\\&
\times \prod_{i\neq j}(z_{i}-z_{j})^{-\eta n_{i}}
 \prod_{i,j}(w_i-z_j)^{p_in_j} 
\\& \propto  \delta^\prime_n \ \mathcal{I}_\alpha 
\prod_{i<j}(z_{i}-z_{j})^{qn_{i}n_{j}}	
\prod_{i\neq j}(z_{i}-z_{j})^{-\eta n_{i}}
\prod_{i,j}(w_i-z_j)^{p_in_j}
\end{split}
\end{equation}
where $\delta_n^\prime = 1$ iff the total number of particles $M = (\eta N - \mathcal{P} - \sum_{k=1}^{Q}p_k)/q$ and $\delta_n^\prime = 0$ otherwise. Particularly we take $\mathcal{P} = -1$ and $\mathcal{P} = -2$ for the cases of two and four quasiholes and $\mathcal{P} = 0$ for the case without quasiholes. Therefore, with $\eta =1$ we achieve the number of particles $M = \frac{N}{2}$ for all the cases with and without quasiholes in the states. The nonzero charge at infinity leads to edge effects.

The results are presented in Fig.\ \ref{Fig_1} and Fig.\ \ref{Fig_2} on a lattice of size $N = 112$ for two and four quasiholes respectively with their different positions. The values of the lattice densities are given by the colorbar with an errorbar of size $\sim 10^{-4}$ arising from the Monte Carlo simulation. It is perceptible that the quasiholes are localized, screened well and the density profile varies with the distance from the quasiholes. Fig.\ \ref{Fig_1}$(b,c)$ and Fig.\ \ref{Fig_2}$(b,c)$ illustrate that there is no singularity in the wavefunction when the quasiholes approach the lattice sites. It only increases the probability that the corresponding lattice sites are unoccupied. 

The radius of an Ising quasihole in the continuum was established in Ref \onlinecite{BAB14} by considering the second moment \cite{ZL4,RNB1}  of the excess charge distribution, and the result was $2.8 l_0$, where $l_0$ is the magnetic length. We plot this number in Fig.\ \ref{Fig_1}$(a)$ and in Fig.\ \ref{Fig_2}$(a)$ for comparison. It is seen that the size of the quasihole in the lattice is comparable to the size in the continuum. Similar results were found for Laughlin quasiholes in Ref \onlinecite{ZL4}.

\textit{Charge. -} In the fractional quantum Hall effect, if we take the charge of the fermionic particles to be $ -1$ then the quasiholes are expected to carry charge $\frac{p_k}{q}$. Now,  the Ising quasiholes in the Moore-Read states in the continuum for $q = 2$ carry an amount of charge $ 0.25$. Experimental measurements of the quasihole charges are in Ref.\ \onlinecite{Others10,Others5,Others13,AS8}. It is thereby indispensable to investigate if they fetch similar charge in the lattice models also. Let us work out the excess charge of the $k$th quasihole elucidated to be the sum of minus the density profile $\rho(z_i)$ over a circular region of radius $r$ around the quasihole \cite{Anne4}

\begin{equation}\label{Eq_14}
Q_{k}(w_{k}) = - \sum_{[i \in \{1,2,....,N\}\big| |z_i - w_{k}|\leq r] } \rho(z_i)
\end{equation}
where $k \in \{1,2,....,Q\}$ and $\rho(z_i)$ is defined in \eqref{Eq_13}. The charge of the quasihole is defined as the value that the total excess charge converges to for large $r$, provided the region is far from the edge and also far from any other quasiholes in the system.

We use the data of Fig.\ \ref{Fig_1}$(a)$ and Fig.\ \ref{Fig_2}$(a)$ to compute the excess charges and plot it in Fig.\ \ref{Fig_1}$(d)$ and Fig.\ \ref{Fig_2}$(d)$ which ensures that with the increment of the radial distances from the quasiholes,  the charges approach the value of $\simeq 0.25$ upto some ignorable uncertainties of order $\sim 10^{-4}$ coming from the Monte Carlo simulation for both the cases of two and four quasiholes concurrently.

\section{Quasihole braiding statistics} \label{Sec.Anyon_braiding}

 \begin{figure*}
 	\includegraphics[width=0.5\textwidth]{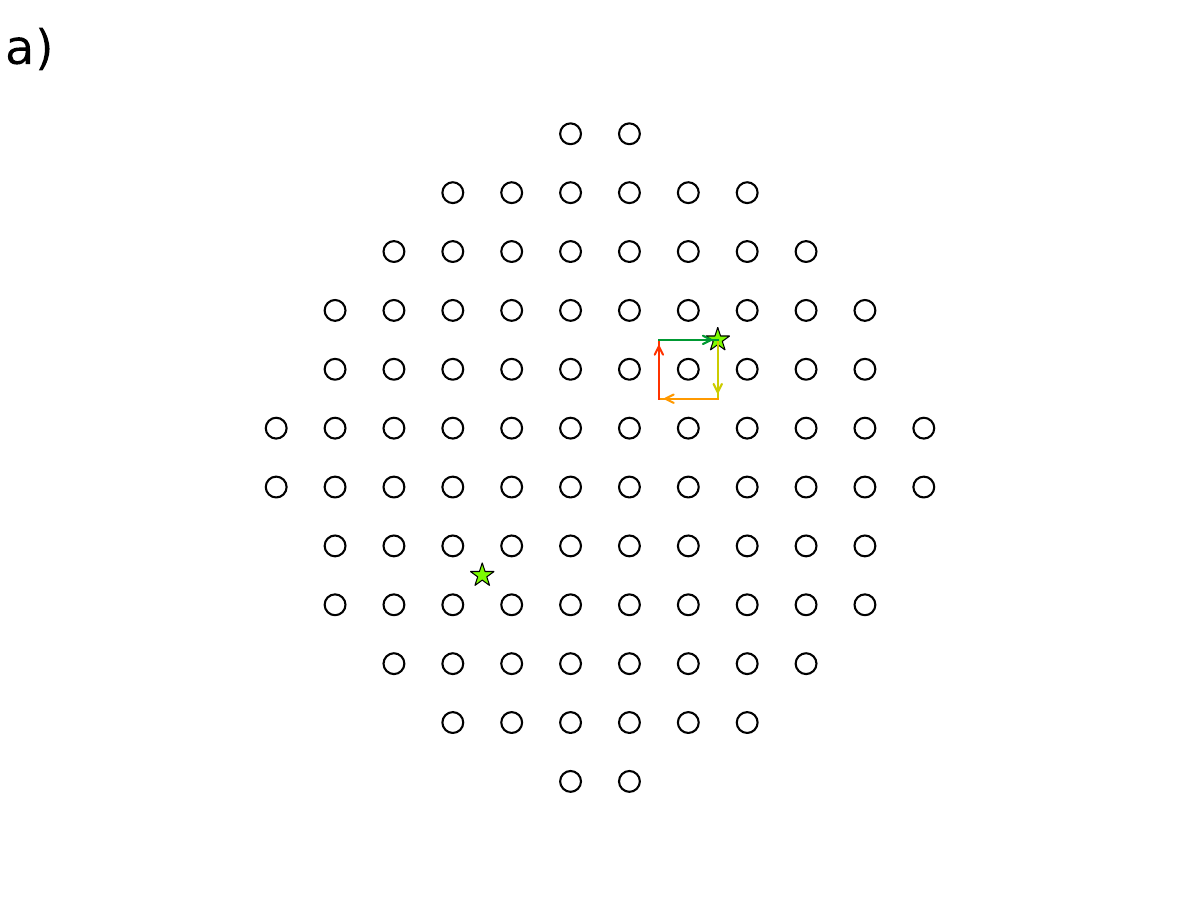}\hfill
 	\includegraphics[width=0.5\textwidth]{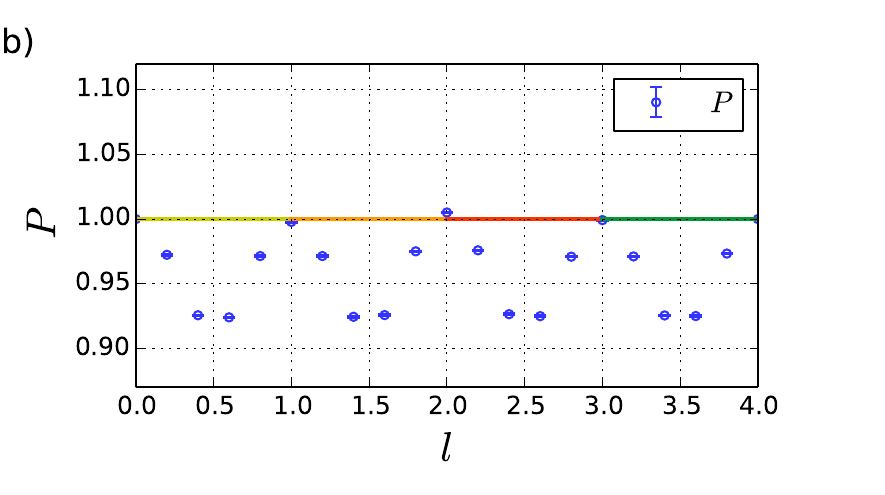}
 	\caption{In $(a)$ circles denote the lattice sites and stars denote the quasiholes. We move one quasihole around one lattice site through a closed loop along the path midway in the lattice plaquette while keeping the others fixed. We choose a lattice of size $N = 96$ and place the quasiholes in the bulk and sufficiently separated from each other. In $(b)$ the inverse ratio between the overlaps at its $l$ th and initial $(l = 0)$ positions i.e.\ $P = \frac {C_0^2}{C_l^2}$ is plotted as a function of the different moves i.e.\ $l$ of the circulating quasihole. It shows that the norm of the conformal block varies with the period of the lattice (upto some numerical uncertainty arising from the simulation and finite size effects).}\label{Fig_3}
 \end{figure*}

\begin{figure*}
	\includegraphics[width=0.5\textwidth]{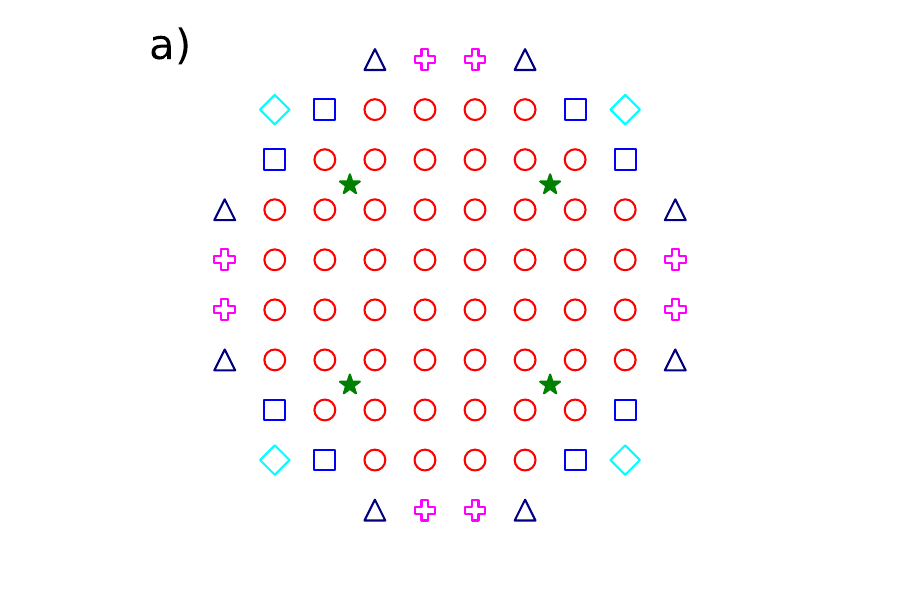}\hfill
	\includegraphics[width=0.5\textwidth]{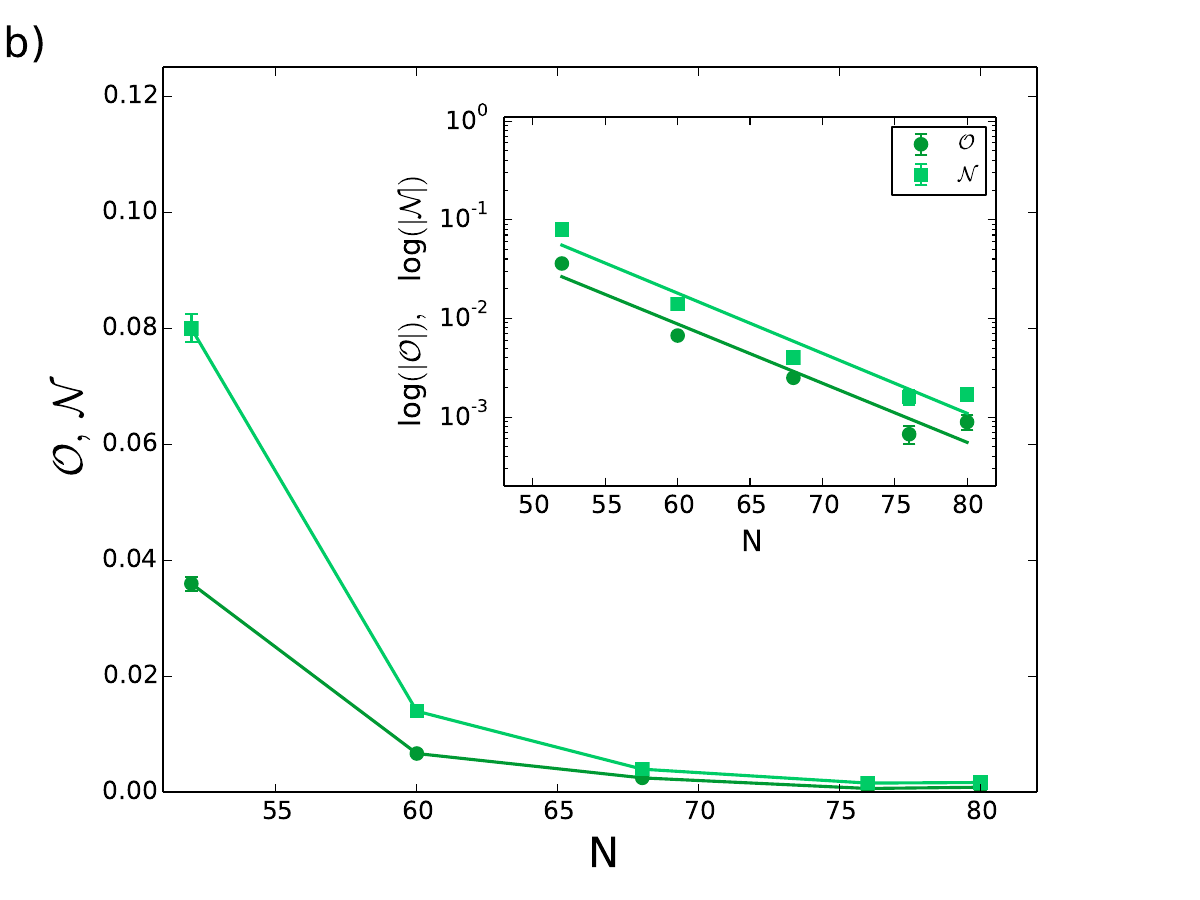}\hfill
	\caption{We keep the quasiholes (stars) fixed in the bulk and sufficiently separated from each other and increase the lattice size by putting more lattice sites as shown in $(a)$. Here circles corresponds to $N = 52$ to begin with and then we increase the lattice size for $N = 60$ (squares), $N = 68$ (pluses), $N = 76$ (triangles) and $N = 80$ (diamonds). We plot in $(b)$ the variations of the overlaps as $\mathcal{O}$ (circles) and $\mathcal{N} $ (squares) respectively as a function of the lattice size. The inset shows the data in the semi log scale. Results depict that the quantities of interest in both the plots are following an exponential decay for sufficiently large lattice sizes. We show a linear fitting of the data points in the insets to conclude the variations as  exponential decay (see text). So, it is expected that in the thermodynamic limit $N \longrightarrow \infty$, the states $\Psi_I$ and $\Psi_\psi$ are going to be orthogonal with the same norm (errorbars are small).}\label{Fig_4}  
\end{figure*}

The results in Sec.\ \ref{Sec.Anyon_properties} show that the quasiholes in the system are localized, well screened with radii of a few lattice constants and with charge $\simeq 0.25$. This provides support for claiming those as Ising quasiholes. These license to go for the braiding statistics of the quasiholes.

To compute braiding, we adiabatically circulate one quasihole around another quasihole along a closed path $\Gamma$  (e.g.\ $w_k$ around $w_j$). This compels the normalized state to pick up a phase matrix and hence, it is transfigured as $|\Psi_\alpha\rangle \longrightarrow \gamma_M \gamma_B |\Psi_\alpha\rangle$. Here, we have two contributions in the story namely the monodromy matrix, i.e.\ the phase matrix arising from the analytic continuation properties of the states, which is denoted by $\gamma_M$, and the Berry matrix $\gamma_B = e^{i\theta_B}$ with elements \cite{C.Nayak8}

\begin{equation}\label{Eq_26}
\big[\theta_B\big]_{\alpha\beta} =i\sum_{k=1}^Q\oint_\Gamma \langle\Psi_\alpha|\frac{\partial \Psi_\beta}{\partial w_k}\rangle dw_k
+ c.c.
\end{equation}
Now it has been proved for the case of the continuum by Bonderson et al.\ in Ref.\ \onlinecite{C.Nayak8} that if the conformal blocks of the states \eqref{Eq_5a} exhibit matrix elements, which are independent of the quasihole positions as long as they are well separated and also the matrix is diagonal in the basis specified by the conformal blocks then the Berry matrix becomes trivial i.e.\ proportional to the identity matrix with an Abelian phase factor as the Aharonov-Bohm phase due to the circulation of the quasihole in the background magnetic field. When a particle of charge $q'$ gets circulated in a magnetic field $B$ through a closed loop of area $A$, it picks up a phase factor of $e^{-2 \pi i q' BA/hc}$ known as the Aharonov-Bohm phase \cite{Others4}, where $h$ is the Planck's constant and $c$ is the speed of light in free space. In this scenario, the quasihole braiding statistics can be read off directly from the analytic continuation alone. 

Now, we investigate the aforesaid conditions for the case of lattice systems. We inscribe the Berry matrix elements to circulate the $k$th quasihole as  
\begin{equation}\label{Berry1}
\begin{split}
\big[\theta_B\big]_{\alpha\beta} &=i\oint_\Gamma \langle\Psi_\alpha|\frac{\partial \Psi_\beta}{\partial w_k}\rangle dw_k
+c.c.
\end{split}
\end{equation}
We use $|\Psi_\alpha \rangle = \frac{1}{C_\alpha}\sum_{n} \Psi_\alpha|n\rangle$ and $|\Psi_\beta \rangle = \frac{1}{C_\beta}\sum_{n^\prime} \Psi_\beta|n^\prime\rangle$ with $\langle n|n^\prime \rangle = \delta_{nn^\prime}$ to write
\begin{equation}\label{Berry2}
\begin{split}
& \langle\Psi_\alpha|\frac{\partial \Psi_\beta}{\partial w_k}\rangle  = \sum_{n} \frac{ \bar{\Psi}_\alpha}{C_\alpha} \frac{\partial}{\partial w_k} \Big( \frac{\Psi_\beta}{C_\beta} \Big) \\&
= \frac{\partial}{\partial w_k} \Big( \sum_{n} \frac{ \bar{\Psi}_\alpha} {C_\alpha}  \frac{\Psi_\beta}{C_\beta} \Big) - \sum_{n} \frac{\Psi_\beta}{C_\beta}  \frac{\partial}{\partial w_k} \frac{ \bar{\Psi}_\alpha} {C_\alpha}\\&
= \frac{\partial}{\partial w_k} \Big( \frac{1}{C_\alpha C_\beta} \sum_{n} \bar{\Psi}_\alpha \Psi_\beta \Big) - \frac{1}{C_\alpha C_\beta}\sum_{n} \Psi_\beta \frac{\partial \bar{\Psi}_\alpha}{\partial w_k} \\&  \qquad \qquad \qquad  - \frac{1}{C_\beta}\sum_{n} \bar{\Psi}_\alpha\Psi_\beta \frac{\partial}{\partial w_k}\Big( \frac{1}{C_\alpha} \Big)
\end{split}
\end{equation}
As per our definition the wavefunctions are normalized. If we show that the wavefunctions are orthogonal i.e.\ $\langle \Psi_\alpha | \Psi_\beta \rangle = \delta_{\alpha \beta}$ then we can write Eq \eqref{Berry2} as 
\begin{equation}\label{Berry3}
\begin{split}
 \langle\Psi_\alpha|\frac{\partial \Psi_\beta}{\partial w_k}\rangle  & = \frac{\partial}{\partial w_k}\delta_{\alpha\beta} - \frac{1}{C_\alpha C_\beta}\sum_{n} \Psi_\beta \frac{\partial \bar{\Psi}_\alpha}{\partial w_k} \\&  \qquad \qquad \qquad  - C_\alpha\delta_{\alpha\beta} \frac{\partial}{\partial w_k}\Big( \frac{1}{C_\alpha} \Big)\\&
= - C_\alpha\delta_{\alpha\beta} \frac{\partial}{\partial w_k}\Big( \frac{1}{C_\alpha} \Big)
\end{split}
\end{equation}
since $\bar{\Psi}_\alpha$ is independent of $w_k$ (it only depends on $\bar{w}_k$). Then we can write the Berry matrix elements as 
\begin{equation}\label{Berry4}
\begin{split}
\big[\theta_B\big]_{\alpha\beta} &= -i \oint_\Gamma  C_\alpha\delta_{\alpha\beta} \frac{\partial}{\partial w_k}\Big( \frac{1}{C_\alpha} \Big) dw_k + c.c. \\&
= i \delta_{\alpha\beta}  \oint_\Gamma \frac{1}{C_\alpha} \Big( \frac{\partial C_\alpha}{\partial w_k} \Big) dw_k + c.c. \\&
= i \delta_{\alpha\beta}  \oint_\Gamma I dw_k + c.c.
\end{split}
\end{equation}
where $I =  \frac{1}{C_\alpha} \Big( \frac{\partial C_\alpha}{\partial w_k} \Big) = \frac{\partial \ln(C_\alpha)}{\partial w_k}$. Now, if $C_\alpha$ (hence $\ln(C_\alpha)$) is periodic in $w_k$ then we have that $\gamma_B$ is equal to the identity matrix i.e.\  $\big[\gamma_B\big]_{\alpha\beta} = \delta_{\alpha \beta}$. Under this circumstance, the quasihole braiding statistics can be evaluated directly from the analytic continuation alone. For the braiding properties to be the same as in the continuum, we further need that $C_\alpha$ and $C_\beta$ are the same. Therefore we have  two sufficient conditions as 
 
 \begin{enumerate}
\item[(i)]$|\sum_{n_i} \Psi^*_\alpha \Psi_{\beta}| = \mathcal{C} \delta_{\alpha \beta} $ up to exponentially small finite  size effects and $ \mathcal{C}  $ is a constant, and
 \item[(ii)]  $C_\alpha$ is periodic when we move one quasihole through a closed loop
 \end{enumerate}

Let us study the braiding statistics extensively for two and four quasihole cases by moving the $k$th quasihole around the $j$th one adiabatically through a closed path.\\

\subsection{Two quasiholes scenario}
 Below we evaluate the Berry matrix and the monodromy matrix in details.\\

\textit{Berry matrix :} We have a single generator of the conformal fields and hence only one state from \eqref{Eq_5a}. Then the Berry matrix \eqref{Eq_26} emerges to be only a phase. We investigate the variation of $C^2$ with the quasihole coordinates while placing them in the bulk and isolated from each other. Henceforth, we keep one quasihole (let us pick up the $k$th one with $k \in \{1,2\}$ symbolizing the quasiholes) moving around one lattice site through a closed loop while keeping the other quasiholes fixed. We choose the path to be along the midway in the lattice plaquette as pictured in Fig \ref{Fig_3}$(a)$ and we expect the same result to hold if we move the quasihole through any other path as well. We inspect the inverse ratio between the overlaps at its $l$ th and initial $(l = 0)$ positions as a function of the different moves i.e.\ $l$ of the circulating quasihole. We denote this ratio as 

\begin{equation}\label{2qh_moves}
P = \frac{C_0^2}{C_l^2}
\end{equation}
We compute $P$ and find the periodic variation of $C^2$ with different positions of the moving quasihole as displayed in Fig \ref{Fig_3}$(b)$. This indeed satisfies the condition (ii) above (and since we have only one wavefunction here condition (i) is not needed). We use Metropolis Monte Carlo simulation to achieve quite large system sizes. Detailed analysis for the technique used is explained in Appendix \ref{AppC}.
In this case as we pointed out earlier the Berry phase contribution is given by $\gamma_B = 1$.\\

\textit{Monodromy matrix :} Now, the counter-clockwise exchange of the two quasiholes gives rise to the monodromy matrix which is just a phase factor here. This analytic continuation can be obtained straightforwardly from the state \eqref{Eq_5a} at face value and it leads to the statistical phase $\gamma_M = e^{i \pi \big[ \frac{p_jp_k}{q} - \frac{1}{8} \big]}$. Also the counter-clockwise circulation of the quasihole around the lattice sites gives rise to the phase $e^{- 2 \pi i p_k/q}$ which can be interpreted as the Aharonov-Bohm phase of a particle with charge $p_k/q$ circulating around a closed loop which encloses the background magnetic flux (taking the standard particle charge $= -1$)\\

Investigations of the braiding properties above lead to the fact that the exchange of two quasiholes gives rise to a phase factor only. This means the quasiholes here abide by Abelian braid statistics.

\subsection{Four quasiholes scenario}

 \begin{figure*}
 	\includegraphics[width=0.33\textwidth]{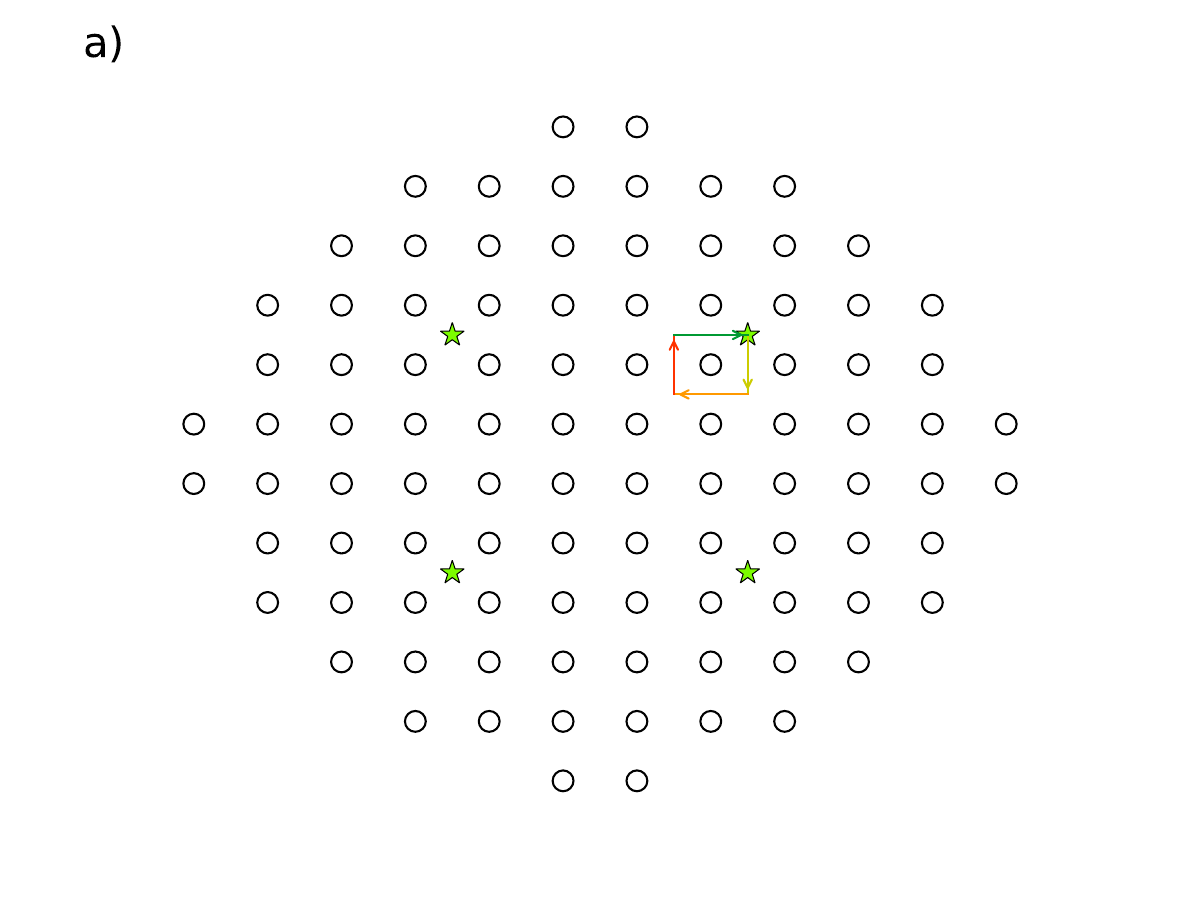}\hfill
 	\includegraphics[width=0.33\textwidth]{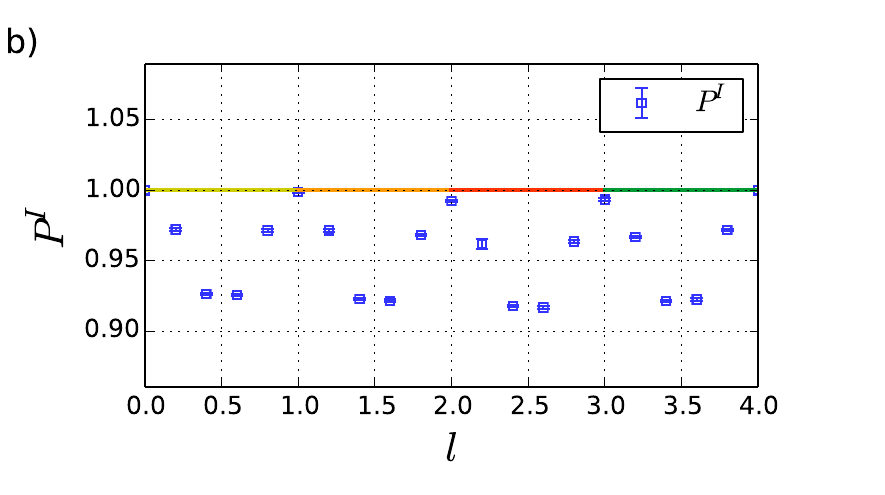}\hfill
 	\includegraphics[width=0.33\textwidth]{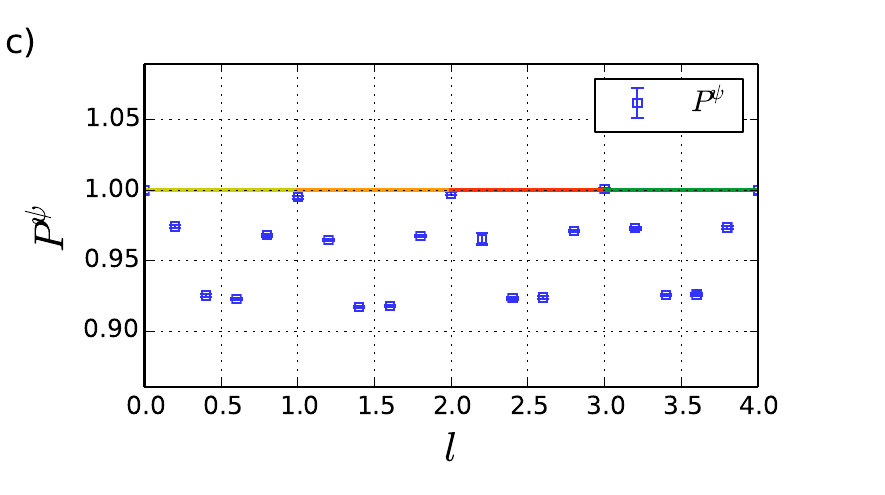}
 	\caption{In $(a)$ circles denote the lattice sites and stars denote the quasiholes. We place the quasiholes in the bulk and sufficiently separated from each other. We move one quasihole around one lattice site through a closed loop along the path midway in the lattice plaquette while keeping the other quasiholes fixed and we choose a lattice of size $N = 96$.  We plot the inverse ratio between the overlaps at its $l$ th and initial $(l = 0)$ positions for both the wavefunctions i.e.\ $ P^I =  \frac{C_{I0}^2}{C_{Il}^2}$ and $ P^\psi =  \frac{C_{\psi 0}^2}{C_{\psi l}^2}$ (marked by squares) as a function of the different moves i.e.\ $l$ of the circulating quasihole in $(b)$ $\&$ $(c)$ respectively.  It shows that the norm of the conformal block varies with the period of the lattice (upto some numerical uncertainty arising from the simulation and finite size effects).}\label{Fig_5}  
 \end{figure*}

Let us proceed to study the Berry matrix and the monodromy matrix extensively in this case.\\

\textit{Berry matrix :}  We have two conformal blocks giving rise to two degenerate states denoted by $\Psi_I$ and $\Psi_\psi$. In this case, the Berry matrix elements are given by \eqref{Berry4} with $ \alpha,\beta \in \{I, \psi\}$. Now, we compute the overlap matrix between the states and we utilize Metropolis Monte Carlo simulations to acquire quite large system sizes. We study now the condition (i). We denote the quantities by $\mathcal{O}$ and $\mathcal{N}$ respectively as

\begin{equation}\label{overlap1}
\mathcal{O} =  \frac{| \sum_{n_i}  \Psi_I^{*}  \Psi_\psi| }{\sqrt{\sum_{n_i} |\Psi_I|^2  \sum_{n_i} |\Psi_\psi|^2}}
\end{equation}
and
\begin{equation}\label{overlap2}
\mathcal{N} = 1 - \frac{\sum_{n_i} |\Psi_I|^2}{\sum_{n_i} |\Psi_\psi|^2}
\end{equation}
We keep the quasiholes fixed and sufficiently separated from each other and increase the lattice size by putting more lattice sites as shown in \ref{Fig_4}$(a)$. We plot the aforementioned quantities in Fig \ref{Fig_4}$(b)$ as a function of the lattice size. Detailed analysis for the technique used is discussed in Appendix \ref{AppC}. Fig \ref{Fig_4}$(b)$ depicts that the quantities of interests follow an exponential decay for sufficiently large lattice sizes. In the inset, we show a linear fit of the data points to conclude the variations are as  $e^{-\lambda N}$ with a decay factor of $\lambda = 0.0599$ and $0.0608$ for $\mathcal{O}$ and $\mathcal{N}$ concurrently. So, in the thermodynamic limit $N \longrightarrow \infty$, the states are expected to be orthogonal with the same norm (upto some numerical uncertainty arising from the simulation). This study license us to note down $|\sum_{n_i} \Psi^*_\alpha \Psi_{\beta}| = \mathcal{C} \delta_{\alpha \beta} + \mathcal{O}(e^{-\lambda N})$ where $\mathcal{C}$ is a constant and $\mathcal{O}(e^{-\lambda N})$ is an exponentially decaying factor of the system size and in the thermodynamic limit, this factor is vanished. Henceforth, the overlap matrix becomes the identity matrix.

Now, to research how $C_\alpha^2$ behaves with the quasihole positions we use the same formalism as in the case of two quasiholes.  We probe here the inverse ratio between the overlaps at the $l$ th and initial ($l = 0$) positions of the moving quasihole (let us choose $k$th one with $k \in \{1,2,3,4\}$ symbolizing the quasiholes) as a function of its different moves i.e.\ $l$ for both the states i.e.\ $\alpha,\beta \in \{I,\psi\}$ by keeping the other quasiholes fixed. Since the overlap matrix becomes diagonal for sufficiently large $N$, it is enough to investigate here only for the diagonal elements i.e.\ $\alpha = \beta$. We denote this ratio as

\begin{equation}\label{4qh_moves}
P^\alpha = \frac{C_{\alpha 0}^2}{C_{\alpha l}^2}
\end{equation}
with $\alpha \in \{I,\psi\}$. We compute $P^\alpha$ and found the periodic variation of $C_\alpha^2$ for both the states with different positions of the moving quasihole. This indeed satisfies the condition (ii). The results are presented in Fig \ref{Fig_5}$(b)$ and $(c)$ for the states $\Psi_I$ and $\Psi_\psi$ respectively affirming that $C_\alpha^2$ varies with the period of the lattice (upto some numerical uncertainty arising from the simulation and finite size effects). We move here the quasihole along the path midway between the lattice sites and we expect the same to hold if we move the quasihole through any other path as well. We show here the result for the circulation of one quasihole. We check that the same happens if we do similar investigation for the other quasiholes as well.

Under this circumstances as we mentioned before the Berry matrix contribution is given by $\gamma_B = \hat{I}$ where $\hat{I}$ is the identity matrix.\\

\textit{Monodromy matrix :} Now, let us investigate the analytic continuation of the states \eqref{Eq_5a} at face value and thereby compute the monodromy matrix $\gamma_M$. We choose the $j$th and $k$th quasiholes (with $j,k \in \{1,2,3,4\}$ symbolizing the quasiholes) to be exchanged in the counter-clockwise fashion while keeping the others fixed. The states $\Psi_I$ and $\Psi_\psi$ are transformed under this exchange $w_j \rightleftarrows w_k$ as \cite{C.Nayak8} follows\\
For $w_1 \rightleftarrows w_2$ or equivalently $w_3 \rightleftarrows w_4$ :
\begin{equation}\label{Eq_29}
\begin{split}
& \Psi_I \mapsto  e^{i \pi \big[ \frac{p_jp_k}{q} - \frac{1}{8} \big]} \Psi_I \\&
\Psi_\psi \mapsto  e^{i \pi \big[ \frac{p_jp_k}{q} - \frac{1}{8} \big]} i\Psi_\psi : \quad j = 1(3), k = 2(4)
\end{split}
\end{equation}
For $w_2 \rightleftarrows w_3$ or equivalently $w_1 \rightleftarrows w_4$ :
\begin{equation}\label{Eq_30}
\begin{split}
& \Psi_I \mapsto  e^{i \pi \big[ \frac{p_jp_k}{q} + \frac{1}{8} \big]} \frac{\Psi_I - i \Psi_\psi}{\sqrt{2}}\\&
\Psi_\psi \mapsto  e^{i \pi \big[ \frac{p_jp_k}{q} + \frac{1}{8} \big]} \frac{-i\Psi_I + \Psi_\psi}{\sqrt{2}} : \quad j = 2(1), k = 3(4)
\end{split}
\end{equation}
For $w_1 \rightleftarrows w_3$ or equivalently $w_2 \rightleftarrows w_4$ :
\begin{equation}\label{Eq_31}
\begin{split}
& \Psi_I \mapsto  e^{i \pi \big[ \frac{p_jp_k}{q} + \frac{1}{8} \big]} \frac{\Psi_I + \Psi_\psi}{\sqrt{2}}\\&
\Psi_\psi \mapsto  e^{i \pi \big[ \frac{p_jp_k}{q} + \frac{1}{8} \big]} \frac{-\Psi_I + \Psi_\psi}{\sqrt{2}} : \quad j = 1(2), k = 3(4)
\end{split}
\end{equation}
where the $\rightleftarrows$ symbol is used to denote the exchange of the quasiholes in the counter-clockwise fashion. Exploitation of Eq \eqref{Eq_29} - \eqref{Eq_31} allows to inscribe the monodromy matrix under the analytic continuation to transmute the states $[\Psi_I,\Psi_\psi]^T \mapsto \gamma_M^{j \rightleftarrows k} [\Psi_I,\Psi_\psi]^T$ as
\begin{equation}\label{Eq_32}
\begin{split}
& \gamma_M^{1 \rightleftarrows 2 / 3 \rightleftarrows 4} =  e^{i \pi \big[ \frac{p_jp_k}{q} - \frac{1}{8} \big]} 
\begin{bmatrix}
1&0\\
0&i
\end{bmatrix}\\&
\gamma_M^{2 \rightleftarrows 3 / 1 \rightleftarrows 4} =  e^{i \pi \big[ \frac{p_jp_k}{q} + \frac{1}{8} \big]} \frac{1}{\sqrt{2}}
\begin{bmatrix}
1&-i\\
-i&1
\end{bmatrix}\\&
\gamma_M^{1 \rightleftarrows 3 / 2 \rightleftarrows 4} =  e^{i \pi \big[ \frac{p_jp_k}{q} + \frac{1}{8} \big]} \frac{1}{\sqrt{2}}
\begin{bmatrix}
1&-1\\
1&1
\end{bmatrix}
\end{split}
\end{equation}\\
Also the counter-clockwise circulation of the quasihole around the lattice sites gives rise to the phase $e^{- 2 \pi i p_k/q}$ which can be interpreted as the Aharonov-Bohm phase of a particle with charge $p_k/q$ circulating around a closed loop which encloses the background magnetic flux (taking the standard particle charge $= -1$). It is seen in Eq \eqref{Eq_32} that the monodromy matrices are the same as found in the continuum \cite{C.Nayak8}. Also they do not commute with each other and hence serve themselves as the members of the braid group. Consequently, investigation of the exchange operations in Eq \eqref{Eq_29} - \eqref{Eq_31} or coequally the matrices in Eq \eqref{Eq_32} comprise the building blocks of the non-Abelian Braid statistics of the states \eqref{Eq_5a} with four quasiholes and thereby affirming the quasiholes here as of non-Abelian nature.\\

 \section{Parent Hamiltonians}\label{Sec.Hamiltonians}
 We have constructed till now Moore-Read wavefunctions hosting two and four quasiholes in lattice systems. Naturally, it is interesting to investigate whether the states in \eqref{Eq_5a} could be defined as the ground states of some Hamiltonians defined on the lattice.
 
  Several works towards this direction have been done recently, for example, in Ref  \onlinecite{Anne5} the Hamiltonian was proposed for the lattice Laughlin state containing quasiholes, and in Ref \onlinecite{Anne4}, Hamiltonians for the lattice Moore-Read state without quasiholes were introduced. In this paper, we fill up the gap by incorporating quasiholes in lattice Moore-Read states and evaluating the Hamiltonian $\forall \ q \geq 2$ accommodating an even number $Q$ of quasiholes. We start by computing the Hamiltonians for $\eta = 1$ and afterwards generalize to the $\eta < 1$ case.\\

 \subsection{Construction for $\eta = 1$}
 
 We take $\eta = 1$ in \eqref{Eq_1} for the lattice limit. The CFT states in \eqref{Eq_5a} are constructed from conformal field correlators which can be utilized to derive the parent Hamiltonians by using null fields of the considered CFT. Null fields have the property that when inserted in conformal field correlators of primary fields, the expectation value becomes zero \cite{Anne4, Anne5}. Explicit derivation of the null fields are done in appendix A.
Now, following the methodology used in Ref \onlinecite{Anne4} we use the null fields to derive in appendix B that the following $q$ operators
   
   \begin{equation}\label{Eq_20} 
   \begin{split}
   \Lambda^{0} &=\sum_{i} d_i, 
   \end{split}
   \end{equation}  
   \begin{equation}\label{Eq_21} 
   \begin{split}
   \underset{p=1,\ldots,q-2}{\Lambda_i^{p}} &=\sum_{j(\neq i)} \frac{1}{(z_i-z_j)^{p}} d_j n_i,
   \end{split}
   \end{equation}
   \begin{equation}\label{Eq_22} 
   \begin{split}
   \Lambda_i^{q-1} = & \sum_{j(\neq i)} \frac{d_j n_i}{(z_i-z_j)^{q}}  + \sum_{j(\neq i)}\sum_{h(\neq i)} \frac{[qn_j-1]d_hn_i}{(z_i-z_h)^{q-1}(z_i-z_j)}\\&
   +  \sum_{j} \sum_{h(\neq i)} \frac{p_jd_hn_i}{(z_i-z_h)^{q-1}(z_i-w_j)}
   \end{split}
   \end{equation}   
annihilate the wavefunction in \eqref{Eq_5a}, i.e.\ $\Lambda_i^a |\Psi_\alpha \rangle = 0$. Here $ d_j$ is defined to be the hardcore bosonic/fermionic annihilation operators for $q$ odd/even acting on the lattice site $j$. The total number of particles at the $j$th lattice site is $n_j =  d^\dagger_jd_j$.  Explicitly, these operators can be written in the matrix form with respect to the basis $(|0\rangle, |1\rangle)$ acting on the $j$th lattice site as
 \begin{align}
 d_j= & \mathcal{S}\begin{bmatrix}
 0&1\\
 0&0
 \end{bmatrix},\ \ \ \
 d_j^\dagger=\mathcal{S}\begin{bmatrix}
 0&0\\
 1&0
 \end{bmatrix},\nonumber
\quad n_j=&\begin{bmatrix}
 0&0\\
 0&1
 \end{bmatrix},\ \ \ \
 \end{align}
 where $\mathcal{S} = (-1)^{(q+1)\sum_{k=1}^{j-1} n_k}$ is the sign factor.
 
 We have $\Lambda_i^a|\Psi_\alpha\rangle  = 0, \quad a\in \{0,1,....,q-1\}$. It follows that the Hermitian operator
 
 \begin{equation}\label{Eq_24} 
 H=\sum_{i=1}^N \sum_{a=0}^{q-1} \Lambda_i^{a\dagger}\Lambda_i^{a}
 \end{equation}
 is a parent Hamiltonian for the state in \eqref{Eq_5a}. 
 
 \subsection{Construction for $\eta < 1$} 
 
 Following the procedure used in Ref \onlinecite{Anne2}  and using the annihilation operators derived in Eq \eqref{Eq_20} - \eqref{Eq_22} we derive here the parent Hamiltonians for the wavefunction for $\eta < 1$ by placing appropriate charges at infinity. We note that the wavefunction with $\eta < 1$ (let us denote it $|\Psi_\alpha^\eta\rangle$) has the number of particles as $M = (\eta N - \sum_{k=1}^{Q}p_k)/q$ and from Eq \eqref{Eq_15} we have it as $M = (N - \mathcal{P} - \sum_{k=1}^{Q}p_k)/q$ for the wavefunction with $\eta = 1$ and a charge $\mathcal{P}$ at infinity. Therefore we have the same number of particles for both the wavefunctions for the particular choice of $\mathcal{P} = N(1-\eta)$. In the Appendix we have found the allowed values of $\mathcal{P}$ as $\mathcal{P} > \Big(-2q - \sum_{k=1}^{Q}p_k + Q\Big)$. This, with the choice of $\mathcal{P} = N(1-\eta)$, leads to the restriction on $\eta$ as follows
 \begin{equation}
 	\eta < 1 + \frac{1}{N} \Big( 2q + \sum_{k=1}^{Q}p_k - Q \Big)
 \end{equation}
 Therefore, in the thermodynamic limit $N \rightarrow \infty$ the parent Hamiltonians, provided below, are valid for $\eta \leq 1$.

It is to be noted that the wavefunction $|\Psi_\alpha^\eta\rangle$ differs from $|\Psi_\alpha^1\rangle$ by a factor of $\prod_{j \neq l} (z_l - z_j)^{(\eta - 1)n_l}$. Let us introduce the operator $\Theta$ as

\begin{equation}\label{theta1} 
\begin{split}
\Theta  = \prod_{l}\Bigg(  \prod_{j(\neq l)} (z_l-z_j)^{(\eta-1)}  \Bigg)^{n_l}
= \prod_{l}\gamma_l^{n_l} 
\end{split}
\end{equation}  
where we define $\gamma_l = \prod_{j(\neq l)} (z_l-z_j)^{(\eta-1)}$ and hence we have $\Theta |\Psi_\alpha^\eta\rangle = |\Psi_\alpha^1\rangle$. Now, we have $\Lambda_i^a|\Psi_\alpha^1\rangle  = 0$ and it immediately follows that $\Theta^{-1} \Lambda_i^a \Theta |\Psi_\alpha^\eta\rangle = 0, \quad a\in \{0,1,....,q-1\}$. Let us note that

\begin{equation}\label{theta2} 
\begin{split}
\Theta^{-1} d_i \Theta = \prod_{l} \gamma_l^{-n_l} d_i  \prod_{m} \gamma_m^{n_m} 
= \gamma_i^{-n_i} d_i  \gamma_i^{n_i}
= \gamma_i d_i
\end{split}
\end{equation}  
Using \eqref{Eq_20}-\eqref{Eq_22} and \eqref{theta2} we construct the following operators $\Lambda^{'a}_i = \Theta^{-1} \Lambda_i^a \Theta$ as

   \begin{equation}\label{L1} 
   \begin{split}
   \Lambda^{'0} &=\sum_{i} \gamma_i d_i, 
   \end{split}
   \end{equation}  
   \begin{equation}\label{L2} 
   \begin{split}
   \underset{p=1,\ldots,q-2}{\Lambda_i^{'p}} &=\sum_{j(\neq i)} \frac{1}{(z_i-z_j)^{p}} \gamma_j d_j n_i,
   \end{split}
   \end{equation}
   \begin{equation}\label{L3} 
   \begin{split}
   \Lambda_i^{'q-1}  & =\sum_{j(\neq i)} \frac{\gamma_j d_j n_i}{(z_i-z_j)^{q}}  + \sum_{j(\neq i)}\sum_{h(\neq i)}  \frac{[qn_j-1]\gamma_h d_hn_i}{(z_i-z_h)^{q-1}(z_i-z_j)}\\&
   +  \sum_{j} \sum_{h(\neq i)} \frac{p_j\gamma_h d_hn_i}{(z_i-z_h)^{q-1}(z_i-w_j)}
   \end{split}
   \end{equation}   
 Finally, for a fixed number of particles $M = (\eta N - \sum_{k=1}^{Q}p_k)/q$ the positive semi-definite Hermitian operator (parent Hamiltonians) which annihilates the wavefunction for $\eta < 1$ becomes

\begin{equation}\label{Ham} 
H=\sum_{i=1}^N \sum_{a=0}^{q-1} \Lambda_i^{'a\dagger}\Lambda_i^{'a}
\end{equation}

The states in Eq \eqref{Eq_5a} are ground states of $H$ by construction, but the above derivation does not exclude that other states could also be ground states. We have tested numerically for the states with two and four quasiholes and different values of $q$ that the ground state degeneracy in the sector with $M$ particles is, indeed, $1$ and $2$, respectively.\\
 
 The Parent Hamiltonian we derived is long ranged and contains up to five-body terms. In addition to its interest as an exact Hamiltonian, it can be used as a test case for numerical techniques. The Hamiltonian is challenging to implement experimentally, but it may be a starting point for finding simpler Hamiltonian with practically the same ground state physics \cite{Anne4, Others50}.\\

\section{Discussion \& Conclusion}\label{Sec.conclusion}

Analytical models are of great importance to study strongly correlated quantum many-body systems. In this work we constructed arbitrarily sized fractional quantum Hall lattice models containing quasiholes. We comprehensively derived bosonic and fermionic strongly correlated lattice Moore-Read Pfaffian states supporting an arbitrary even number of quasiholes for this lattice model by exploiting conformal field correlators of the underlying Ising CFT. Our construction allows to make an interpolation between lattice models and the continuum via a parameter $\eta$ introduced in the states. 

We investigated the relevant properties like density profile, charge and braiding statistics of the quasiholes by using Metropolis Monte Carlo simulations for $q = 2$. The outcomes displayed that the quasiholes are localized, well screened with radii of a few lattice constants and contain a charge of $ \simeq 0.25$ which agrees with the Ising quasiholes in the continuum. We then probe the topological properties of the states directly by analyzing the fractional braiding statistics of the quasiholes. The investigations show that the two quasiholes behave as if they are Abelian and the four quasihole case ensures the non-Abelian nature of the Ising quasiholes.

By using null fields of the underlying Ising CFT we constructed parent Hamiltonians for $\eta \leq 1$ and $\forall \ q \geq 2$ containing an even number of quasiholes and spanning  the degenerate space.

Due to extreme complexity of the strongly correlated electronic systems, investigation of various fascinating phenomena, for example topology, becomes easier if we have models with analytical ground states. The findings of this article represent CFT and Monte Carlo techique as powerful tools in this direction. Also, analysis and claim in the context of non-Abelian quasiholes are of particular importance regarding topological quantum computation. 

The methodology used to construct the lattice model here is quite general and it would be very interesting to construct and inspect other fractional quantum Hall lattice models containing Abelian and non-Abelian quasiholes e.g.\ Fibonacci quasiholes in $\mathbb{Z}_3$ Read-Rezayi states.\\

\section{Acknowledgments}

We thank Blazej Jaworowski. The authors would like to thank Ignacio Cirac for discussions. JW thanks NSF DMR 1306897 and NSF DMR 1056536 for partial support. GS has been supported by the Spanish grants FIS2015-69167-C2-1-P, QUITEMAD+ S2013/ICE-2801 and SEV-2016-0597. AEBN has been supported by the Carlsberg Foundation and the Villum Foundation during early stages of this project.

%%%%%%%%%%%%%%%%%%%%%%%%%%%%%%%%%%
\allowdisplaybreaks
\appendix
\section{Null fields of the underlying Ising CFT}\label{AppA}
In this appendix, following Ref \onlinecite{Anne4}, we derive that the fields defined in \eqref{Null1}-\eqref{Null3} below are null fields. From the $c = 1$ massless bosonic CFT with compactification radius $\sqrt{q}$ we can define operators \cite{Anne5} as two chiral currents $G^\pm (z) = \ : \psi(z) e^{\pm i \sqrt{q}\phi(z)} :$ and the $U(1)$ conformal current $J(z) = \frac{i}{\sqrt{q}}\partial_z\phi(z)$. 

Then we introduce $q+1$ fields as follows
\begin{equation}\label{Null1} 
\begin{split}
{\chi^{p}(v)} &=\oint_{v}\frac{dz}{2\pi i}\frac{1}{(z-v)^{p}} G^{+}(z) V_{1}(v),
\end{split}
\end{equation}
\begin{equation}\label{Null2} 
\begin{split}
\chi^{q-1}(v) =\oint_{v}\frac{dz}{2\pi i} &\left[\frac{1}{(z-v)^{q-1}} G^{+}(z) V_{1}(v)\right.\\
& \left.-\frac{1}{(z-v)}V_{2}(v)\right],
\end{split}
\end{equation}  
\begin{equation}\label{Null3} 
\begin{split}
\chi^{q}(v) &=\oint_{v}\frac{dz}{2\pi i}\frac{1}{z-v} \left[\frac{1}{(z-v)^{q-1}}G^{+}(z)V_{1}(v)\right] \\
&- \oint_{v}\frac{dz}{2\pi i}\frac{1}{z-v} qJ(z)V_{2}(v)
\end{split}
\end{equation}     
where we define $\mathcal{V}_{n_j}(v) = \chi_{n_j}(v)V_{n_j}(v)$ from Eq \eqref{Eq_1} - \eqref{Eq_2}. Eq \eqref{Null1} represents $q-1$ fields since $p \in \{0,1,....,q-3,q-2\}$.

We explicitly derive that the fields in Eq \eqref{Null1} - \eqref{Null3} are null fields. We do the similar calculation for our case as done in Ref \onlinecite{Anne4}. The CFT states are obtained from the operators $\mathcal{V}_{n_j}(z_j)$ and $W_{p_j}(w_j)$ as defined in Eq \eqref{Eq_1} - \eqref{Eq_3} and we here allow occupancy $n_j \in \{0,1,2\}$ rather than just $n_j \in \{0,1\}$. We here consider the lattice limit $\eta = 1$. We need to use the following expressions \cite{Anne4}
\begin{equation}\label{Eq_A1}
:e^{i\alpha\phi(z)}::e^{i\beta\phi(v)}: \  = (z-v)^{\alpha\beta}:e^{i\alpha\phi(z)+i\beta\phi(v)}:,
\end{equation} 
\begin{equation}\label{Eq_A2}
\psi_i(z)\psi_j(v) = \delta_{ij}\bigg[\frac{1}{z-v} + (z-v)A(v)+....\bigg],
\end{equation} 
\begin{equation}\label{Eq_A3}
\begin{split}
e^{i\phi(z)} &\simeq e^{i[\phi(v) + (z-v)\partial_v\phi(v)]}\\& = e^{i\phi(v)}e^{i(z-v)\partial_v\phi(v)}\\& \simeq e^{i\phi(v)} [1 + i(z-v)\partial_v\phi(v)],
\end{split}
\end{equation}
\begin{equation}\label{Eq_A4}
\partial_z\phi(z) = \partial_v\phi(v) + (z-v)\partial^2_v\phi(v) + .... 
\end{equation}  
where $....$ stands for terms that are proportional to $(z-v)^k$ with $k \geq 2$. The particular form of $A(v)$ is not required as we keep in mind that the non-zero contributions of the integrals in the null fields come from the terms having simple poles. The following proofs are applicable for all $q\geq 2$.

\subsection{Null field $\chi^q(v)$}\label{AppA1}
We write $\forall \ q\geq2$
\begin{equation}\label{Eq_A5}
\begin{split}
\chi^{q}(v) &=\oint_{v}\frac{dz}{2\pi i}\frac{1}{(z-v)^{q}}G^{+}(z)V_{1}(v)
\\ & - \oint_{v}\frac{dz}{2\pi i}\frac{1}{z-v} qJ(z)V_{2}(v)\\
&=\mathcal{I}^q_{1}(v) - \mathcal{I}^q_{2}(v)
\end{split}
\end{equation}
where the integration contour is a circle around $v$ and we consider the counter-clockwise direction as the positive one per convention. Now, writing the terms explicitly and using Eq \eqref{Eq_A1} - \eqref{Eq_A4}, we find the non-zero contributions as
\begin{equation}\label{Eq_A6}
\begin{split}
\mathcal{I}^q_{1}(v) & = \oint_{v}\frac{dz}{2\pi i}\frac{1}{z-v}\left[\frac{1}{(z-v)^{q-1}}G^{+}(z)V_{1}(v)\right]\\&
=\oint_{v}\frac{dz}{2\pi i}\frac{1}{z-v}\left[\frac{1}{(z-v)^{q-1}}\psi(z)\psi(v)\right.\\
& \left. \qquad \qquad \qquad \times e^{+ i\sqrt{q}\phi (z)}e^{i(q-1)\phi (v)/\sqrt{q}}\right]\\&
=\oint_{v}\frac{dz}{2\pi i}\frac{1}{z-v}\left[\frac{(z-v)^{q-1}}{(z-v)^{q-1}}\psi(z)\psi(v)\right.\\
& \left. \qquad \qquad \qquad \times e^{i\sqrt{q}\phi (z)+i(q-1)\phi (v)/\sqrt{q}}\right]\\&
=\oint_{v}\frac{dz}{2\pi i}\frac{1}{z-v}\left[\left(\frac{1}{z-v}+(z-v)A(v)+...\right)\right.\\
& \left. \qquad \qquad \qquad \times e^{i\sqrt{q}\phi (z)+i(q-1)\phi (v)/\sqrt{q}}\right]\\&
=\oint_{v}\frac{dz}{2\pi i}\left[\frac{1}{(z-v)^2}e^{i\sqrt{q}\phi (z)+i(q-1)\phi (v)/\sqrt{q}}\right]\\&
=\oint_{v}\frac{dz}{2\pi i} \frac{1}{z-v} \left[i \sqrt{q} \partial_v \phi (v) e^{i(2q-1)\phi (v)/\sqrt{q}}\right]
\end{split}
\end{equation}
and
\begin{equation}\label{Eq_A7}
\begin{split}
\mathcal{I}^q_{2}(v) & = \oint_{v}\frac{dz}{2\pi i}\frac{1}{z-v}\left[qJ(z)V_{2}(v)\right]\\
&=\oint_{v}\frac{dz}{2\pi i}\frac{1}{z-v}\left[\sqrt{q} i \partial_v \phi (z) e^{i(2q-1)\phi (v)/\sqrt{q}}\right]\\
&=\oint_{v}\frac{dz}{2\pi i}\frac{1}{z-v}\left[\sqrt{q} i \partial_v \phi (v) e^{i(2q-1)\phi (v)/\sqrt{q}}\right]\\ 
\end{split}
\end{equation}  
It is seen that, $\mathcal{I}^q_{1}(v) = \mathcal{I}^q_{2}(v)$ which ensures $\chi^q(v)$ as a null field.

\subsection{Null field $\chi^{q-1}(v)$}\label{AppA2}
We write in this case also $\forall \ q\geq2$
\begin{equation}\label{Eq_A8}
\begin{split}
\chi^{q-1}(v) &=\oint_{v}\frac{dz}{2\pi i} \frac{1}{(z-v)^{q-1}}G^{+}(z)V_{1}(v)
\\ & - \oint_{v}\frac{dz}{2\pi i}\frac{1}{z-v} V_{2}(v)\\
&=\mathcal{I}^{q-1}_{1}(v) - \mathcal{I}^{q-1}_{2}(v)
\end{split}
\end{equation}
Now, proceeding in the same way as before we get
\begin{equation}\label{Eq_A9}
\begin{split}
\mathcal{I}^{q-1}_{1}(v) & = \oint_{v}\frac{dz}{2\pi i}\frac{1}{(z-v)^{q-1}}G^{+}(z)V_{1}(v)\\&
=\oint_{v}\frac{dz}{2\pi i}\left[\frac{1}{(z-v)^{q-1}}\psi(z)\psi(v)\right.\\
& \left. \qquad \qquad \qquad \times e^{+ i\sqrt{q}\phi (z)}e^{i(q-1)\phi (v)/\sqrt{q}}\right]\\&
=\oint_{v}\frac{dz}{2\pi i}\left[\frac{(z-v)^{q-1}}{(z-v)^{q-1}}\psi(z)\psi(v)\right.\\
& \left. \qquad \qquad \qquad \times e^{i\sqrt{q}\phi (z)+i(q-1)\phi (v)/\sqrt{q}}\right]\\&
=\oint_{v}\frac{dz}{2\pi i}\left[\left(\frac{1}{z-v}+(z-v)A(v)+...\right)\right.\\
& \left. \qquad \qquad \qquad \times e^{i\sqrt{q}\phi (z)+i(q-1)\phi (v)/\sqrt{q}}\right]\\&
=\oint_{v}\frac{dz}{2\pi i}\left[\frac{1}{(z-v)}e^{i\sqrt{q}\phi (v)+i(q-1)\phi (v)/\sqrt{q}}\right]\\&
=\oint_{v}\frac{dz}{2\pi i}\left[\frac{1}{(z-v)}e^{i(2q-1)\phi(v)/\sqrt{q}}\right] 
\end{split}
\end{equation}   
and
\begin{equation}\label{Eq_A10}
\begin{split}
\mathcal{I}^{q-1}_{2}(v) & = \oint_{v}\frac{dz}{2\pi i}\frac{1}{z-v}V_{2}(v)\\
& =\oint_{v}\frac{dz}{2\pi i}\left[\frac{1}{(z-v)}e^{i(2q-1)\phi(v)/\sqrt{q}}\right]
\end{split}
\end{equation}  
It is seen that, $\mathcal{I}^{q-1}_{1}(v) = \mathcal{I}^{q-1}_{2}(v)$ which ensures $\chi^{q-1}(v)$ as a null field.

\subsection{Null fields ${\chi^{p}(v)}, p \in\{0,1,....,q-2\}$}\label{AppA3}
These null fields are defined $\forall \ q \geq 2$ and $p \in\{0,1,....,q-2\}$. We write
\begin{equation}\label{Eq_A11}
\begin{split}
& {\chi^{p}(v)}  = \oint_{v}\frac{dz}{2\pi i}\frac{1}{(z-v)^{p}}G^{+}(z)V_{1}(v)\\&
=\oint_{v}\frac{dz}{2\pi i}\left[\frac{1}{(z-v)^{p}}\psi(z)\psi(v)\right.\\
& \left. \qquad \qquad \qquad \times e^{+ i\sqrt{q}\phi (z)}e^{i(q-1)\phi (v)/\sqrt{q}}\right]\\&
=\oint_{v}\frac{dz}{2\pi i}\left[\frac{(z-v)^{q-1}}{(z-v)^{p}}\psi(z)\psi(v)\right.\\
& \left. \qquad \qquad \qquad \times e^{i\sqrt{q}\phi (z)+i(q-1)\phi (v)/\sqrt{q}}\right]\\&
=\oint_{v}\frac{dz}{2\pi i}  \frac{(z-v)^{q-1}}{(z-v)^{p}}\left[\left(\frac{1}{z-v}+(z-v)A(v)+...\right)\right.\\
& \left. \qquad \qquad \qquad \times e^{i\sqrt{q}\phi (z)+i(q-1)\phi (v)/\sqrt{q}}\right]\\&
=\oint_{v}\frac{dz}{2\pi i}  \frac{(z-v)^{q-1}}{(z-v)^{p}}\left[\left(\frac{1}{z-v}+(z-v)A(v)+...\right)\right.\\
& \left. \quad \times e^{i(2q - 1)\phi(v)/\sqrt{q}}[1+i\sqrt{q}(z-v)\partial_v\phi(v)+....]\right] \\& = 0
\end{split}
\end{equation}
No term in the above integral has a simple pole to provide a non-zero contribution since, $p \in\{0,1,....,q-2\}$ and thereby ensuring $ {\chi^{p}(v)}$ as null fields.\\

 \section{Operators annihilating the CFT wavefunctions containing an even number of quasiholes}\label{AppB}
 
In the following subsections we derive a set of operators annihilating the wave functions when $q \geq 2$ and $\eta = 1$ for lattice systems with occupancy $n_j \in \{0,1,2\}$ and containing an even number of quasiholes. Next we use these results to derive the same for the lattice systems with occupancy $n_j\in \{0,1\}$ as given in Eq \eqref{Eq_20} - \eqref{Eq_22}. Finally we compute the condition on $\eta$ as mentioned in Sec.\ \ref{Sec.Hamiltonians}B.

We note that if we insert the null fields to the vacuum expectation value of the primary chiral conformal fields in \eqref{Eq_5a}, it leads to the decoupling equations as,
$\langle 0|\prod_{k=1}^{Q}W(w_k)\prod_{j=1}^{i-1}\mathcal{V}_{n_j}(z_j)\chi^a(z_i)\prod_{j=i+1}^{N}\mathcal{V}_{n_j}(z_j)|0\rangle = 0$. The next step is to rewrite these equations in the form $\Lambda_i^a |\Psi^\alpha\rangle = 0$, where $\Lambda_i^a$ are the operators which annihilate the wavefunction. Finally, the Hamiltonian is defined as $H = \sum_{a,i} \Lambda^{a \dagger}_i \Lambda_i^a$.\\

\subsection{$\eta = 1$ and occupancy $n_j \in \{0,1,2\}$}\label{AppB1}
 
To construct parent Hamiltonians from null fields we note that the correlator vanishes if the field at site $i$ is replaced by a null field. Next we derive decoupling equations satisfied by the CFT correlator in \eqref{Eq_5a} for an even number of quasiholes by deforming the integration contour over the complex plane, moving the operators ($G^+(z)$ and $J(z)$) in the null fields at different positions and using operator product expansions together with the commutation relations as below \cite{Anne4,Anne5}
 \begin{equation}\label{Eq_B1}
 \begin{split}
 G^+(z)\mathcal{V}_{n_j}(z_j) \sim &(-1)^{(j-1)} \left[ \frac{\delta_{n_j,0} \delta_{n'_j,1}}{z-z_j} \right] \mathcal{V}_{n'_j}(z_j),
 \end{split}
 \end{equation}
 \begin{equation}\label{Eq_B2}
 \begin{split}
 G^+(z)W(w_j) \sim 0,
 \end{split}
 \end{equation}
 \begin{equation}\label{Eq_B3}
 \begin{split}
 \mathcal{V}_{n_j}(z_j)G^{+}(z)=(-1)^{(q+1)n_j-1}G^{+}(z)\mathcal{V}_{n_j}(z_j),
 \end{split}
 \end{equation}
 \begin{equation}\label{Eq_B4}
 \begin{split}
 J(z)\mathcal{V}_{n_j}(z_j) \sim \frac{1}{q}\frac{(qn_j - 1)}{z-z_j}\mathcal{V}_{n_j}(z_j),
 \end{split}
 \end{equation}
 \begin{equation}\label{Eq_B5}
 \begin{split}
 J(z)W(w_j) \sim \frac{1}{q}\frac{p_j}{z-w_j}W_{p_j}(w_j),
 \end{split}
 \end{equation}
 \begin{equation}\label{Eq_B6}
 \begin{split}
 :e^{i\alpha\phi(z)}::e^{i\beta\phi(z_j)}: \  = (z-z_j)^{\alpha\beta}:e^{i\alpha\phi(z)+i\beta\phi(z_j)}:,
 \end{split}
 \end{equation}
 \begin{equation}\label{Eq_B7}
 \begin{split}
 :e^{i\alpha\phi(z)}::e^{i\beta\phi(z_j)}:{}=
 (-1)^{\alpha\beta}:e^{i\beta\phi(z_j)}::e^{i\alpha\phi(z)}:,
 \end{split}
 \end{equation}
 \begin{equation}\label{Eq_B8}
 \begin{split}
 \psi_i(z)\psi_j(z_j) = \delta_{ij}(-1)^{n_j}\psi_i(z_j)\psi_j(z)
 \end{split}
 \end{equation}
 where $\sim$ means that we have considered the operator product expansion up to the terms which would give non zero contribution in our results. The total number of particles at the $j$th lattice site is $n_j = n_j^{(1)} + 2n_j^{(2)}$ where $n_j^{(1)} = d^\dagger_jd_j$ and $n_j^{(2)} = d_j'^\dagger d_j'$ define individual number of particles for the two levels $ |0\rangle \leftrightarrow |1\rangle$ and $ |1\rangle \leftrightarrow |2\rangle$ respectively. Those operators acting on the states of the three level system lead to the following equations with proper sign factor as \cite{Anne4}
 \begin{align}\label{Eq_B9-B12}
 d_j |{n_j}\rangle&=(-1)^{(q+1)\sum_{k=1}^{j-1} n_k} \begin{cases}0 & n_j=0 \\
 |{0}\rangle & n_j=1\\
 0 & n_j=2
 \end{cases}\\
 d_j^\dagger |{n_j}\rangle&=(-1)^{(q+1)\sum_{k=1}^{j-1} n_k}\begin{cases}|{1} \rangle& n_j=0 \\
 0 & n_j=1\\
 0 & n_j=2
 \end{cases}\\
 d_j' |{n_j}\rangle&=(-1)^{(q+1)\sum_{k=1}^{j-1} n_k}\begin{cases}0 & n_j=0 \\
 0 & n_j=1\\
 |{1}\rangle & n_j=2
 \end{cases}\\
 d_j'^\dagger |{n_j}\rangle&=(-1)^{(q+1)\sum_{k=1}^{j-1} n_k}\begin{cases}0 & n_j=0 \\
 |{2}\rangle & n_j=1\\
 0 & n_j=2
 \end{cases}
 \end{align}
  Explicitly, the above mentioned operators can be written in the matrix form with respect to the basis $(|0\rangle, |1\rangle, |2\rangle)$ acting on the $j$th lattice site as
 \begin{align}\label{Eq_B13}
 d_j= & \mathcal{S}\begin{bmatrix}
 0&1&0\\
 0&0&0\\
 0&0&0
 \end{bmatrix},\ \ \ \
 d_j^\dagger=\mathcal{S}\begin{bmatrix}
 0&0&0\\
 1&0&0\\
 0&0&0
 \end{bmatrix},\nonumber\\
 d'_j= & \mathcal{S}\begin{bmatrix}
 0&0&0\\
 0&0&1\\
 0&0&0
 \end{bmatrix},\ \ \ \
 d_j'^\dagger=\mathcal{S}\begin{bmatrix}
 0&0&0\\
 0&0&0\\
 0&1&0
 \end{bmatrix},\nonumber\\
n_j^{(1)}=&\begin{bmatrix}
  0&0&0\\
  0&1&0\\
  0&0&0
  \end{bmatrix},\ \ \ \
  n_j^{(2)}=\begin{bmatrix}
  0&0&0\\
  0&0&0\\
  0&0&1
  \end{bmatrix}
 \end{align}
where $\mathcal{S} = (-1)^{(q+1)\sum_{k=1}^{j-1} n_k}$ is the sign factor already defined before. 
 
  \begin{widetext}
 Here, we evaluate the annihilation operator for the CFT wavefunction in detail for the null field $\chi^q(v), \forall \ q \geq 2$. Therefore,\\\\
 
 	\begin{equation}\label{Eq_B14}
 	\begin{split}
 	0 & =  	\langle W(w_1)....W(w_Q)\mathcal{V}_{n_1}(z_1)....\mathcal{V}_{n_{i-1}}(z_{i-1})\chi^q(z_i) \mathcal{V}_{n_{i+1}}(z_{i+1})....\mathcal{V}_{n_N}(z_N)\rangle \\&
 	= \oint_{z_i}\frac{dz}{2\pi i}\frac{1}{(z-z_i)^q}
 	\langle W(w_1)\dots W(w_Q)\mathcal{V}_{n_{1}}(z_{1})\ldots G^{+}(z)V_{1}(z_i) \ldots \mathcal{V}_{n_{N}}(z_{N})\rangle\\&
 	-q\oint_{z_i}\frac{dz}{2\pi i}\frac{1}{z-z_i}\langle W(w_1)\dots W(w_Q)
 	\mathcal{V}_{n_{1}}(z_{1})\ldots J(z)V_2(z_i) \ldots \mathcal{V}_{n_{N}}(z_{N})\rangle \\&
 	= I_1^q + I_2^q 
 	\end{split}
 	\end{equation}
 	
 	The term $I_1^q$ evaluates to :
 	
 	\begin{equation}\label{Eq_B15}
 	\begin{split}
 	&\oint_{z_i}\frac{dz}{2\pi i}\frac{1}{(z-z_i)^q}
 	\langle W(w_1)\dots W(w_Q)\mathcal{V}_{n_{1}}(z_{1})\ldots G^{+}(z)V_{1}(z_i) \ldots \mathcal{V}_{n_{N}}(z_{N})\rangle\\&
 	=-\sum_{j=1(\neq i)}^{N}\oint_{z_j}\frac{dz}{2\pi i}\frac{1}{(z-z_i)^q} \langle W(w_1)\dots W(w_Q)\mathcal{V}_{n_{1}}(z_{1})\ldots  G^{+}(z)V_{1}(z_i) \ldots \mathcal{V}_{n_{N}}(z_{N})\rangle\\&
 	=-(-1)^{i-1}\sum_{j=1}^{i-1}\oint_{z_j}\frac{dz}{2\pi i}\frac{(-1)^{(q+1)\sum_{k=j}^{i-1}n_k}}{(z-z_i)^q}
 	\frac{\delta_{n_j,0} \delta_{n'_j,1}}{z-z_j}
 	\langle W(w_1)\dots W(w_Q) \mathcal{V}_{n_{1}}(z_{1})\ldots \mathcal{V}_{n'_j}(z_j) \ldots V_{1}(z_i) \ldots \mathcal{V}_{n_{N}}(z_{N})\rangle\\
 	&\phantom{=}-(-1)^{i-1}\sum_{j=i+1}^N\oint_{z_j}\frac{dz}{2\pi i}\frac{(-1)^{(q+1)}(-1)^{(q+1)\sum_{k=i+1}^{j-1}n_k}}{(z-z_i)^q}
 	\frac{\delta_{n_j,0} \delta_{n'_j,1}}{z-z_j} \langle W(w_1)\dots W(w_Q)\mathcal{V}_{n_{1}}(z_{1})\ldots\\&\qquad\qquad\quad\qquad\qquad\qquad\qquad\qquad\qquad\qquad\qquad\qquad\dots
 	V_{1}(z_i) \dots \mathcal{V}_{n'_j}(z_j) \ldots \mathcal{V}_{n_{N}}(z_{N})\rangle\\&
 	=-(-1)^{i-1}\sum_{j=1}^{i-1}\frac{(-1)^{(q+1)\sum_{k=j}^{i-1}n_k}}{(z_j-z_i)^q}
 	\delta_{n_j,0} \delta_{n'_j,1} \langle W(w_1)\dots W(w_Q) \mathcal{V}_{n_{1}}(z_{1})\ldots \mathcal{V}_{n'_j}(z_j) \ldots  V_{1}(z_i) \ldots \mathcal{V}_{n_{N}}(z_{N})\rangle\\
 	&\phantom{=}-(-1)^{i-1}\sum_{j=i+1}^{N}\frac{(-1)^{(q+1)}(-1)^{(q+1)\sum_{k=i+1}^{j-1}n_k}}{(z_j-z_i)^q}
\delta_{n_j,0} \delta_{n'_j,1}
 	\langle W(w_1)\dots W(w_Q)\mathcal{V}_{n_{1}}(z_{1})\ldots \\&\qquad\qquad\qquad\qquad\qquad\qquad\qquad\qquad\qquad\qquad\qquad \ldots V_{1}(z_i)\dots \mathcal{V}_{n'_j}(z_j) \ldots \mathcal{V}_{n_{N}}(z_{N})\rangle
 	\\&=-(-1)^{i-1}\sum_{j=1}^{i-1}\sum_{n'_j}\frac{(-1)^{(q+1)\sum_{k=j}^{i-1}n_k}}{(z_j-z_i)^q}
  \delta_{n_j,0} \delta_{n'_j,1} 
 	\langle W(w_1)\dots W(w_Q) \mathcal{V}_{n_{1}}(z_{1})\ldots\mathcal{V}_{n'_j}(z_j) \dots V_{1}(z_i) \ldots \mathcal{V}_{n_{N}}(z_{N})\rangle\\
 	&\phantom{=}-(-1)^{i-1}\sum_{j=i+1}^{N}\sum_{n'_j}\frac{(-1)^{(q+1)}(-1)^{(q+1)\sum_{k=i+1}^{j-1}n_k}}{(z_j-z_i)^q}
  \delta_{n_j,0} \delta_{n'_j,1}
 	\langle W(w_1)\dots W(w_Q)\mathcal{V}_{n_{1}}(z_{1})\ldots \\&\qquad\qquad\qquad\qquad\qquad\qquad\qquad\qquad\qquad\qquad\qquad\qquad\qquad \ldots V_{1}(z_i)\dots \mathcal{V}_{n'_j}(z_j) \ldots \mathcal{V}_{n_{N}}(z_{N})\rangle\\&
 	=-\sum_{j=1}^{i-1}\frac{(-1)^{(q+1)\sum_{k=j+1}^{i-1}n_k}}{(z_j-z_i)^q} \delta_{n_j,0}
 	\Psi_{\alpha}(n_1,\ldots,1,\ldots,1,\ldots,n_N)\\&
 	\phantom{=}-\sum_{j=i+1}^N\frac{(-1)^{(q+1)}(-1)^{(q+1)\sum_{k=i+1}^{j-1}n_k}}{(z_j-z_i)^q} \delta_{n_j,0}
 	\Psi_{\alpha}(n_1,\ldots,1,\ldots,1,\ldots,n_N)\\&
 	=-\sum_{j=1}^{i-1}\frac{(-1)^{(q+1)}(-1)^{(q+1)\sum_{k=j+1}^{i-1}n_k}}{(z_i-z_j)^q} \delta_{n_j,0}
 	\Psi_{\alpha}(n_1,\ldots,1,\ldots,1,\ldots,n_N)\\&
 	\phantom{=}-\sum_{j=i+1}^N\frac{(-1)^{(q+1)\sum_{k=i+1}^{j-1}n_k}}{(z_i-z_j)^q} \delta_{n_j,0}
 	\Psi_{\alpha}(n_1,\ldots,1,\ldots,1,\ldots,n_N)\\&
 	\end{split}
 	\end{equation}
 	To achieve decoupling equations involving the CFT wavefunctions in \eqref{Eq_5a}, we multiply \eqref{Eq_B15} by $|n_1,\ldots,n_{i-1},2,n_{i+1}\ldots,n_N\rangle$ and sum over all $n_k, k \neq i$ and thereby end up with 
 	\begin{equation}\label{Eq_B17}
 	\sum_{j=1(\neq i)}^{N}\frac{1}{(z_i-z_j)^q} d_j d_i'^\dagger|\Psi_{\alpha}\rangle
 	\end{equation}
 	Let us evaluate the term $I_2^q$ :

 	\begin{equation}\label{Eq_B18}
 	\begin{split}
 	&-q\oint_{z_i}\frac{dz}{2\pi i}\frac{1}{z-z_i}\langle W(w_1)\dots W(w_Q)
 	\mathcal{V}_{n_{1}}(z_{1})\dots J(z)V_2(z_i) \ldots \mathcal{V}_{n_{N}}(z_{N})\rangle\\&
 	=q\sum_{j=1(\neq i)}^{N}\oint_{z_j}\frac{dz}{2\pi i}\frac{1}{z-z_i}\langle W(w_1)\dots W(w_Q)
 	\mathcal{V}_{n_{1}}(z_{1})\ldots J(z)V_2(z_i) \ldots \mathcal{V}_{n_{N}}(z_{N})\rangle\\&
 	+ q\sum_{j=1}^{Q}\oint_{w_j}\frac{dz}{2\pi i}\frac{1}{z-z_i}\langle W(w_1)\dots W(w_Q)
 	\mathcal{V}_{n_{1}}(z_{1})\ldots J(z)V_2(z_i) \ldots \mathcal{V}_{n_{N}}(z_{N})\rangle\\&
 	=\sum_{j=1(\neq i)}^{N}\oint_{z_j}\frac{dz}{2\pi i}\frac{1}{z-z_i} \frac{(qn_j-1)}{z-z_j} \langle W(w_1)\dots W(w_Q)
 	\mathcal{V}_{n_{1}}(z_{1})\dots\mathcal{V}_{n_j}(z_j)\dots V_2(z_i) \ldots \mathcal{V}_{n_{N}}(z_{N})\rangle\\&
 	+  \sum_{j=1}^{Q}\oint_{w_j}\frac{dz}{2\pi i}\frac{1}{z-z_i}\frac{p_j}{z-w_j}\langle W(w_1)\dots W(w_Q)
 	\mathcal{V}_{n_{1}}(z_{1})\ldots V_2(z_i) \ldots \mathcal{V}_{n_{N}}(z_{N})\rangle\\&
 	=\sum_{j=1(\neq i)}^{N}\frac{(qn_j-1)}{z_j-z_i}\langle W(w_1)\dots W(w_Q)
 	\mathcal{V}_{n_{1}}(z_{1})\dots\mathcal{V}_{n_j}(z_j)\dots V_2(z_i) \ldots \mathcal{V}_{n_{N}}(z_{N})\rangle\\&
 	+  \sum_{j=1}^{Q}\frac{p_j}{w_j-z_i}\langle W(w_1)\dots W(w_Q)
 	\mathcal{V}_{n_{1}}(z_{1})\ldots V_2(z_i) \ldots \mathcal{V}_{n_{N}}(z_{N})\rangle
 	\end{split}
 	\end{equation}
 \end{widetext}
 Now, we multiply \eqref{Eq_B18} by $|n_1,\ldots,n_{i-1},2,n_{i+1}\ldots,n_N\rangle =\sum_{n'_i}n_i^{(2)}|n_1,\ldots,n'_i,\ldots,n_N\rangle$,
 and sum over all $n_k$, $k\neq i$ to get
 \begin{equation}\label{Eq_B19}
 -\sum_{j=1(\neq i)}^{N}\frac{qn_j-1}{z_i-z_j}n^{(2)}_i|\Psi_{\alpha}\rangle
 -\sum_{j=1}^{Q}\frac{p_j}{z_i-w_j}n^{(2)}_i|\Psi_{\alpha}\rangle
 \end{equation}
 So, summing up \eqref{Eq_B17} and \eqref{Eq_B19}, we achieve finally,
 \begin{equation}\label{Annihilation}
 \begin{split}
\lambda_i^q |\Psi_{\alpha}\rangle = 0
 \end{split}
 \end{equation}
 where 
  \begin{equation}\label{Eq_B20}
  \begin{split}
  & \lambda_i^q = \sum_{j=1(\neq i)}^{N}\frac{d_j d_i'^\dagger}{(z_i-z_j)^q}  -\sum_{j=1(\neq i)}^{N}\frac{qn_j-1}{z_i-z_j}n^{(2)}_i \\& \qquad 
  -\sum_{j=1}^{Q}\frac{p_j}{z_i-w_j}n^{(2)}_i
  \end{split}
  \end{equation}
 Proceeding in the same way and using the other null fields in Eq \eqref{Null1} - \eqref{Null3} in the main text, we end up with the following annihilation operators for the CFT wavefunctions in the spin $1$ case as
 \begin{equation}\label{Eq_B21}
 \begin{split}
 \lambda^{0} &=\sum_{i}d_i, 
 \end{split}
 \end{equation}  
 \begin{equation}\label{Eq_B22}
 \begin{split}
{\lambda_i^{p}} &=\sum_{j(\neq i)} \frac{1}{(z_i-z_j)^{p}} d_j d_i'^\dagger,
 \end{split}
 \end{equation}
 \begin{equation}\label{Eq_B23}
 \begin{split}
 \lambda_i^{q-1} &=\sum_{j(\neq i)}\frac{1}{(z_i-z_j)^{q-1}} d_j d_i'^\dagger+n_i^{(2)}
 \end{split}
 \end{equation}

\subsection{$\eta = 1$ and occupancy $n_j \in \{0,1\}$}\label{AppB2}

Following the procedure used in Ref \onlinecite{Anne4} we derive here operators annihilating the wave function for the occupancy $n_j \in \{0,1\}$ by using the operators derived in Eq \eqref{Eq_B20} - \eqref{Eq_B23}. We divide the Hilbert space $\mathcal{H}_1 + \mathcal{H}_2$ into two subspaces $\mathcal{H}_1$ and $\mathcal{H}_2$. $\mathcal{H}_1$ is the space consisting of all states with no doubly occupied sites, and $\mathcal{H}_2$ is the space consisting of all states with at least one doubly occupied site. Then operators for the occupancy $n_j \in \{0,1\}$ system lie in $\mathcal{H}_1$  and for the occupancy $n_j \in \{0,1,2\}$ system reside in $\mathcal{H}_1 + \mathcal{H}_2$. We project the operators in $\mathcal{H}_1 + \mathcal{H}_2$ to $\mathcal{H}_1$ to get the operators. 

We multiply the operators $\lambda^a_i, \quad a\in \{0,1,....,q\}$ derived in Eq \eqref{Eq_B20} - \eqref{Eq_B23} by $d^{'}_i$ from the left. Since, $d^{'}_i d^{'\dagger}_i = n_i^{(1)}$ we have the operators $d^{'}_i \lambda^a_i, \quad a\in \{0,1,....,q-2\} $ annihilating the wavefunction for the occupancy $n^{(1)}_j \in \{0,1\}$ since these act on $\mathcal{H}_1$ only. It is to be noted that $d^{'}_i \lambda^{q-1}_i$ annihilates the wavefunctions for the occupancy $n_j \in \{0,1,2\}$ and hence we can write 

 \begin{equation}\label{Eq_B24}
 \begin{split}
 \Bigg[d^{'}_i + \sum_{j(\neq i)} \frac{1}{(z_i-z_j)^{q-1}} d_j n_i^{(1)} \Bigg] |\Psi_\alpha \rangle = 0
 \end{split}
 \end{equation} 
which allows us to replace the $ d^{'}_i$ operator in $d^{'}_i \lambda^{q}_i$ by $- \sum_{h(\neq i)} \frac{1}{(z_i-z_h)^{q-1}} d_hn_i^{(1)}$. So, after making the projection the operators become

   \begin{equation}\label{B25} 
   \begin{split}
   \Lambda^{0} &=\sum_{i} d_i, 
   \end{split}
   \end{equation}  
   \begin{equation}\label{B26} 
   \begin{split}
   \underset{p=1,\ldots,q-2}{\Lambda_i^{p}} &=\sum_{j(\neq i)} \frac{1}{(z_i-z_j)^{p}} d_j n_i^{(1)},
   \end{split}
   \end{equation}
   \begin{equation}\label{B27} 
   \begin{split}
   \Lambda_i^{q-1} =\sum_{j(\neq i)} \frac{d_j n_i^{(1)}}{(z_i-z_j)^{q}}  &+ \sum_{j(\neq i)} \sum_{h(\neq i)} \frac{[qn_j^{(1)}-1]d_hn_i^{(1)}}{(z_i-z_h)^{q-1}(z_i-z_j)}\\&
   +  \sum_{j} \sum_{h(\neq i)} \frac{p_jd_hn_i^{(1)}}{(z_i-z_h)^{q-1}(z_i-w_j)}
   \end{split}
   \end{equation}   
These operators all annihilate the occupancy $n^{(1)}_j \in \{0,1\}$ wave function. In the main text we denote $n^{(1)}_j$ as $n_j$.

\subsection{Condition on $\eta$}

We first derive the condition on the charge $\mathcal{P}$ at infinity and thereby using the relation $\mathcal{P} = N(1-\eta)$ we compute the condition on $\eta$. The starting point is that if we insert a null field the correlator becomes zero as

\begin{equation}\label{B28}
	\langle \prod_{k=1}^{Q}W(w_k)  \Xi_\mathcal{P}(\infty) \prod_{j=1}^{i-1} \mathcal{V}_{n_j}(z_j)\chi^a(z_i)\prod_{j=i+1}^{N}\mathcal{V}_{n_j}(z_j)\rangle = 0
\end{equation}
where $a \in \{ 0,1,....,q \}$. Let us derive the above correlator for different parts of the null fields. For the term 
 \begin{equation}
 \begin{split}
- \oint_{v}\frac{dz}{2\pi i}\frac{1}{z-v} qJ(z)V_{2}(v)
 \end{split}
 \end{equation}  
we have
 \begin{equation}\label{B31}
\begin{split}
&-q\oint_{z_i}\frac{dz}{2\pi i}\frac{1}{z-z_i}\langle W(w_1)\dots W(w_Q)  \Xi_\mathcal{P}(\xi)
\mathcal{V}_{n_{1}}(z_{1})\ldots
\\& \qquad \qquad \times
J(z)V_2(z_i) \ldots \mathcal{V}_{n_{N}}(z_{N})\rangle
\end{split}
\end{equation}
Now, we proceed as before and multiply the term in Eq \eqref{B31} by $|n_1,\ldots,n_{i-1},2,n_{i+1}\ldots,n_N\rangle =\sum_{n'_i}n_i^{(2)}|n_1,\ldots,n'_i,\ldots,n_N\rangle$,
and sum over all $n_k$, $k\neq i$ to get

\begin{equation}\label{B32}
\begin{split}
-\sum_{j=1(\neq i)}^{N}\frac{qn_j-1}{z_i-z_j}n^{(2)}_i|\Psi_\alpha^1\rangle
& -\sum_{j=1}^{Q}\frac{p_j}{z_i-w_j}n^{(2)}_i|\Psi_\alpha^1\rangle\\&
-\frac{\mathcal{P}}{z_i - \xi}n^{(2)}_i |\Psi_\alpha^1\rangle
\end{split}
\end{equation}
where the last term in Eq \eqref{B32} vanishes in the limit $\xi \rightarrow \infty$.

Similarly for the term 
 \begin{equation}
 \begin{split}
 - \oint_{v}\frac{dz}{2\pi i}\frac{1}{z-v}V_{2}(v)
 \end{split}
 \end{equation}  
we get 
 \begin{equation}
 \begin{split}
 &-\oint_{z_i}\frac{dz}{2\pi i}\frac{1}{z-z_i}\langle W(w_1)\dots W(w_Q)  \Xi_\mathcal{P}(\xi)
 \mathcal{V}_{n_{1}}(z_{1})\ldots
 \\& \qquad \qquad \times
V_2(z_i) \ldots \mathcal{V}_{n_{N}}(z_{N})\rangle
 \end{split}
 \end{equation}
Again by proceeding in the same way as before we find
\begin{equation}
\begin{split}
n^{(2)}_i|\Psi_\alpha^1\rangle
\end{split}
\end{equation}
So it means that the charge at infinity term does not have any effect on these. Now we consider the following term
 \begin{equation}
 \begin{split}
\oint_{v}\frac{dz}{2\pi i}\frac{1}{(z-v)^a}G^{+}(z)V_{1}(v)
 \end{split}
 \end{equation}   
 where $a \in \{ 0,1,....,q \}$. We have 
 \begin{equation}\label{B30}
 \begin{split}
 &\oint_{z_i}\frac{dz}{2\pi i}\frac{1}{(z-z_i)^a}
 \langle W(w_1)\dots W(w_Q) \Xi_\mathcal{P}(\xi) \mathcal{V}_{n_{1}}(z_{1})\ldots 
 \\& \qquad \qquad \times G^{+}(z)V_{1}(z_i) \ldots \mathcal{V}_{n_{N}}(z_{N})\rangle
 \end{split}
 \end{equation}
The  term in Eq \eqref{B30} after contour deformation becomes

\begin{equation}\label{B33}
\begin{split}
&-\sum_{j=1(\neq i)}^{N}\oint_{z_j}\frac{dz}{2\pi i}\frac{1}{(z-z_i)^q} \langle W(w_1)\dots W(w_Q)
\Xi_\mathcal{P}(\xi)\\& \qquad \qquad \times
\mathcal{V}_{n_{1}}(z_{1})\ldots 
G^{+}(z)V_{1}(z_i) \ldots \mathcal{V}_{n_{N}}(z_{N})\rangle
\\&
-\oint_{\xi}\frac{dz}{2\pi i}\frac{1}{(z-z_i)^q} \langle W(w_1)\dots W(w_Q)
\Xi_\mathcal{P}(\xi)\\& \qquad \qquad \times
\mathcal{V}_{n_{1}}(z_{1})\ldots 
G^{+}(z)V_{1}(z_i) \ldots \mathcal{V}_{n_{N}}(z_{N})\rangle
\end{split}
\end{equation}
We proceed as before and multiply the first term in Eq \eqref{B33} by $|n_1,\ldots,n_{i-1},2,n_{i+1}\ldots,n_N\rangle$ and sum over all $n_k, k \neq i$ and thereby end up with 

\begin{equation}\label{B34}
\sum_{j=1(\neq i)}^{N}\frac{1}{(z_i-z_j)^q} d_j d_i'^\dagger |\Psi_\alpha^1\rangle
\end{equation}

\begin{widetext}	
	Let us evaluate the second term in Eq \eqref{B33} as
	
	\begin{equation}\label{B35}
	\begin{split}
	-(-1)^{i-1+p} \sum_{n_i} \delta_{n_i = 1} \oint_{\xi}  \frac{dz}{2\pi i} \frac{(-1)^{-(i-1)n_i}}{(z-z_i)^a} (-1)^{(q+1)\sum_{k=1}^{i-1}n_k}
	\langle W(w_1)\dots W(w_Q)
	\Xi_\mathcal{P}(\xi)
	\mathcal{V}_{n_{1}}(z_{1})\ldots 
	G^{+}(z)V_{1}(z_i) \ldots \mathcal{V}_{n_{N}}(z_{N})\rangle
	\end{split}
	\end{equation}
	As we know the expression of the correlator,  we can compute the contour integral as
	
	\begin{equation}\label{B36}
	\begin{split}
	&-(-1)^{i-1+\mathcal{P}} \delta_{\mathcal{P}<0} \sum_{n_i} \delta_{n_i = 1} 
	\lim_{z \rightarrow \xi} \frac{1}{(-\mathcal{P}-1)!}\frac{d^{-\mathcal{P}-1}}{dz^{-\mathcal{P}-1}}
	\frac{(-1)^{-(i-1)n_i}}{(z-z_i)^a} (-1)^{(q+1)\sum_{k=1}^{i-1}n_k}
	\delta_n \text{Pf}(A) \prod_{i,j} (w_i - z^{'}_j)^{\frac{-1}{2}}\\&
	\times \prod_{j} (z-z_j)^{(qn_j - 1)} \prod_{j}(-1)^{(j-1)n_j}  \prod_{j} (\xi-z_j)^{(qn_j - 1)\frac{\mathcal{P}}{q}} \prod_{j} (-1)^{(j-1)n_j}
	\prod_{j<k}(z_j - z_k)^{(qn_j - 1)(qn_k -1)/q}\\&
	\prod_{j<k}(w_j-w_k)^{\frac{p_jp_k}{q}} \prod_{j,k}(w_j - z_k)^{(qn_k-1)p_j/q}
	\end{split}
	\end{equation}

\end{widetext}
where $\delta_n = 1$ iff the total number of particles $M = (N - \mathcal{P} - \sum_{k} p_k - q)/q$ and 0 otherwise. Now Eq \eqref{B36}$ \ =0$ gives rise to the condition on the choice of $\mathcal{P}$ and hence $\eta$. This also keeps the derived annihilation operators unchanged. It is to be noted that the expression in Eq \eqref{B36} is zero if $\mathcal{P}>0$ due to the delta factor $\delta_{\mathcal{P}<0} $. By inspecting the derivative and the exponent of the polynomial we find that Eq \eqref{B36} is also zero when $\mathcal{P} > -q -a - \sum_{k}p_k + Q$. Since, $a \in \{ 0,1,....,q \}$ we can safely use the maximum value of $a$ in that expression to write $\mathcal{P} > -2q - \sum_{k}p_k + Q$. By using $\mathcal{P} = N(1-\eta)$ we get immediately the condition on $\eta$ as. 

\begin{equation}
	\eta < 1 + \frac{1}{N} \Big( 2q + \sum_{k=1}^{Q}p_k - Q \Big)
\end{equation}

 In the thermodynamic limit $N \rightarrow \infty$ this condition becomes $\eta < 1$.

\section{Technical details of the overlap computation using the Metropolis Monte Carlo technique}\label{AppC}

Here we display the numerical details of the Metropolis Monte Carlo technique used to derive the overlaps in Sec.\ \ref{Sec.Anyon_braiding} of the main text. We explicitly show here the computation of the overlap 

\begin{equation}\label{C1}
O = \frac{ |\sum_{n_i}\Psi_\alpha^{*}  \Psi_\beta |}{\sqrt{\sum_{n_i}|\Psi_\alpha|^2\sum_{n_i}|\Psi_\beta|^2}}
\end{equation}
with $\alpha,\beta \in \{I, \psi\}, \alpha \neq \beta$ and the evaluation of other overlaps can be done following the same procedure. We first write

\begin{equation}\label{C2}
\begin{split}
 \frac{ |\sum_{n_i}\Psi_\alpha^{*}  \Psi_\beta |}{\sqrt{\sum_{n_i}|\Psi_\alpha|^2\sum_{n_i}|\Psi_\beta|^2}} & 
 = \frac{\Lambda_{\alpha \beta}}{\sqrt{\Omega_{\alpha \beta}\Omega_{\beta \alpha}}}
\end{split}
\end{equation}
where we have

\begin{equation}\label{C3}
\begin{split}
& \Lambda_{\alpha \beta} = \frac{|\sum_{n_i} |\Psi_\alpha \Psi_\beta| \frac{\Psi_\alpha^* \Psi_\beta}{|\Psi_\alpha \Psi_\beta|}|}{\sum_{n_i} |\Psi_\alpha \Psi_\beta|}\\&
\Omega_{\alpha \beta} = \frac{\sum_{n_i} |\Psi_\alpha \Psi_\beta| \frac{|\Psi_\alpha|}{|\Psi_\beta|}}{\sum_{n_i} |\Psi_\alpha \Psi_\beta|}\\&
\Omega_{\beta\alpha } = \frac{\sum_{n_i} |\Psi_\alpha \Psi_\beta| \frac{|\Psi_\beta|}{|\Psi_\alpha|}}{\sum_{n_i} |\Psi_\alpha \Psi_\beta|}
\end{split}
\end{equation}

Now, the quantities $\frac{\Psi_\alpha^* \Psi_\beta }{|\Psi_\alpha \Psi_\beta |}, \frac{|\Psi_\alpha |}{|\Psi_\beta |}$ and $\frac{|\Psi_\beta |}{|\Psi_\alpha |}$ can be obtained by Metropolis Monte Carlo sampling over the lattice occupancy distribution with weight $|\Psi_\alpha \Psi_\beta|$.

\bibliography{bibfile}
\end{document}